\pdfoutput=1
\documentclass[%
 aip,
 amsmath,amssymb,
 preprint,%
]{revtex4-1}
\usepackage{graphicx}%
\usepackage{xcolor}
\usepackage{caption}
\usepackage{subcaption}
\usepackage{booktabs}
\usepackage{multirow}
\usepackage{hyperref}

\usepackage{float}

\newcommand{\MHW}{modified Hasegawa-Wakatani}
\newcommand{\HW}{Hasegawa-Wakatani}
\newcommand{\genmodel}{multichannel flow matching}

\makeatletter
\@booleanfalse\titlepage@sw
\def\frontmatter@abstract@produce{%
  \par
  \addvspace{\frontmatter@preabstractspace}%
  \begingroup
   \dimen@\baselineskip
   \setbox\z@\vtop{\unvcopy\absbox}%
   \advance\dimen@-\ht\z@\advance\dimen@-\prevdepth
   \@ifdim{\dimen@>\z@}{\vskip\dimen@}{}%
  \endgroup
  \begingroup
   \prep@absbox
   \unvbox\absbox
   \post@absbox
  \endgroup
  \@ifx{\@empty\mini@notes}{}{\mini@notes\par}%
  \addvspace\frontmatter@postabstractspace
}%
\makeatother

\begin{document}

\title{A generative foundation model for 2D magnetized plasma turbulence}%

\author{Junyi Guo}
\affiliation{Sibley School of Mechanical and Aerospace Engineering, Cornell University, Ithaca, New York 14853, USA}
\author{Meet H. Parikh}
\affiliation{Sibley School of Mechanical and Aerospace Engineering, Cornell University, Ithaca, New York 14853, USA}
\author{Jian-Xun Wang}
\affiliation{Sibley School of Mechanical and Aerospace Engineering, Cornell University, Ithaca, New York 14853, USA}
\author{Ben Zhu}
\email[]{ben.zhu@columbia.edu}
\affiliation{Department of Applied Physics and Applied Mathematics, Columbia University, New York, NY 10027, USA}
\affiliation{Columbia Fusion Research Center, Columbia University, New York, NY 10027, USA}

\begin{abstract}
We present a proof-of-principle study demonstrating a conditional generative foundation model for self-organizing magnetized plasma turbulence. Built on a multichannel flow-matching framework with a spectral decomposition strategy, the model is trained on direct numerical simulations of 2D modified Hasegawa-Wakatani (MHW) turbulence across the hydrodynamic to adiabatic parameter range. Without retraining, this \textit{single} model successfully performs three distinct downstream tasks: (i) unconditional generation of statistically faithful quasi-steady-state snapshots that recover the directional energy spectra and amplitude statistics, from which statistical transport quantities, including turbulent flux and effective diffusivity, alongside their uncertainties, are inferred directly; (ii) super-resolution from $4\times$ and $8\times$ downsampled observations to reconstruct fine-scale turbulence and large-scale zonal structures; and (iii) full-field inpainting from sparse spatial observations covering as little as $\sim 10\%$ of the domain. The generated fields faithfully reproduce regime-dependent turbulence vs. zonal-flow states and energy partitioning. Under extreme sparsity the long-correlation zonal flows are pinned by only a handful of observations, while the residual uncertainty concentrates in the small-scale turbulent fluctuations. These results establish flow-matching generative models as powerful, unified surrogates for state generation, statistical inference, and sparse-data reconstruction, providing a scalable baseline for 3D plasma turbulence and experimental applications.
\end{abstract}

\maketitle%

\section{Introduction}
\label{sec:intro}
Magnetized plasma turbulence and its nonlinear interaction with self-generated mean flows, in particular zonal flows, govern cross-field transport in essentially every magnetic-confinement fusion device and underlie the formation of transport barriers and confinement transitions~\cite{Diamond2005}. A predictive understanding of this turbulence-flow interplay is therefore essential for fusion reactor design and operation, but is made difficult by the extreme range of spatial and temporal scales, the strongly nonlinear and chaotic character of the dynamics, and the high cost of high-fidelity gyrokinetic or fluid simulations.

For the past four decades, reduced fluid models of drift-wave turbulence have provided a tractable yet physically meaningful setting in which to study these phenomena. The \HW{} (HW) model~\cite{HasegawaWakatani1983} in its original form describes resistive drift-wave turbulence in a magnetized slab through two coupled equations for the perturbed density and the electrostatic potential. In the \emph{modified} \HW{} (MHW) model~\cite{Numata2007}, the resistive coupling is restricted to the non-zonal components of the fields, thus allowing the spontaneous generation of stable zonal flows through turbulent Reynolds stresses and the subsequent suppression of cross-field transport. The MHW model thus retains, within a minimal two-field 2D system, the essence of drift-wave-zonal-flow interaction. Its two asymptotic limits, hydrodynamic at small adiabaticity and quasi-adiabatic at large adiabaticity, bracket two regimes of practical interest: nearly isotropic turbulence and strongly anisotropic turbulence-zonal-flow states~\cite{Numata2007}. Statistical and phase-transition properties of HW/MHW turbulence have continued to attract attention~\cite{Anderson2017,vanderWetering2026}.

Recent years have seen rapid adoption of machine learning (ML) in plasma physics, ranging from data-driven closures~\cite{Ma2020,Maulik2020} and surrogate models~\cite{Zhu2025} to disruption predictors~\cite{KatesHarbeck2019}. The HW/MHW family has served as a test bed for many of these developments, including data-driven rediscovery of the model equations from simulation data~\cite{Abramovic2022}, non-intrusive reduced-order modelling~\cite{Gahr2024}, calibration of coarse-grid dissipation against high-fidelity turbulence statistics~\cite{Li2025}, and neural-operator surrogates that capture the bifurcating drift-wave and zonal-flow dynamics~\cite{vanderWetering2026}. However, conventional supervised learning approaches face  fundamental challenges in turbulence modeling: they are deterministic (unable to sample from the conditional distribution of turbulent states) and task-specific (a model trained for forward prediction cannot perform inference or reconstruction without retraining). Generative models address these limitations by learning the full data distribution $p(\mathbf{x}|\alpha)$ rather than a point map, enabling (i) ensemble sampling with uncertainty quantification, (ii) conditioning on partial observations for inverse problems without task-specific retraining, and (iii) potential generalization across regimes. Generative approaches have begun to enter plasma turbulence research, through GAN-based deconvolution for large-eddy simulations~\cite{Castagna2024} and variational autoencoder-based generative surrogates for long-time transport simulations~\cite{Clavier2025}. In parallel, generative models such as diffusion models~\cite{Ho2020} and the more recent flow-matching framework~\cite{Lipman2023} have rapidly matured into powerful tools for high-fidelity reconstruction, super-resolution and inpainting in hydrodynamic turbulence~\cite{Du2024,Fan2025,Oommen2025,Parikh2026,fan2026generative}. However, these plasma applications remain task-specific and single-regime. A unified conditional generative \emph{foundation} model for magnetized plasma turbulence, one that spans turbulence regimes and serves state synthesis, statistical inference, and sparse-data reconstruction within a single learned prior, and captures the coexistence and competition between self-generated zonal flows and small-scale drift-wave turbulence, has so far been lacking.

In this work, we present a proof-of-principle study of a generative foundation model for magnetized plasma turbulence, using the 2D MHW system as a controlled testbed. We build a single conditional generative model on direct numerical simulations spanning the full hydrodynamic-to-adiabatic range, using a multichannel flow-matching model~\cite{Parikh2026} augmented with a spectral decomposition strategy~\cite{Fan2025} to mitigate spectral bias and capture multiscale
dynamics. The trained model demonstrates three key capabilities: it unconditionally synthesizes
statistically faithful turbulent state realizations that recover regime-dependent energy
spectra and zonal-flow morphology, from which statistical transport quantities such as the turbulent particle flux are inferred directly, and the \emph{same} trained model supports two practically
important conditional reconstruction tasks—physical-state super-resolution from heavily downsampled low-resolution fields, and full-field reconstruction from spatially sparse observations \emph{without any task-specific retraining or finetuning}. In addition, we characterize how reconstruction  uncertainty is distributed across scales under extreme observation sparsity: zonal flows are readily pinned by sparse observations due to long correlation lengths, while higher uncertainty concentrates in small-scale turbulent fluctuations. These results have implications
for plasma diagnostics and data assimilation in more realistic experimental settings.

The remainder of the paper is organized as follows. Section~\ref{sec:model} introduces the MHW equations, outlines the simulation campaign and the construction of the training and evaluation datasets. Section~\ref{sec:flowmatch} describes the multichannel flow-matching generative model used in this study. Section~\ref{sec:results} presents the unconditional-generation and transport-inference results, the super-resolution and inpainting results, and the generalization to adiabaticity values not used in training, together with quantitative assessments of statistical fidelity. Section~\ref{sec:summary} summarizes the findings and discusses open issues and future directions.

\section{Physics models and data generation}
\label{sec:model}

\subsection{The modified Hasegawa--Wakatani system}
\label{sec:mhw}

We consider the 2D MHW model in a doubly periodic slab with Cartesian coordinates $(x,y)$, where $x$ denotes the radial direction and $y$ is the binormal/poloidal direction. Under the standard drift-wave normalization, the model evolves the fluctuating plasma density $n(x,y,t)$ and electrostatic potential $\phi(x,y,t)$ according to
\begin{align}
    \partial_t n + \{\phi, n\}
    &= \alpha\,(\tilde\phi - \tilde n)
       - \kappa\,\partial_y\phi
       - \nu_n\,(-\nabla_\perp^2)^{N} n, \label{eq:mhw_n}\\
    \partial_t \varpi + \{\phi, \varpi\}
    &= \alpha\,(\tilde\phi - \tilde n)
       - \nu_\varpi\,(-\nabla_\perp^2)^{N}\varpi, \label{eq:mhw_w}
\end{align}
where $\varpi \equiv \nabla_\perp^2 \phi$ is the vorticity, $\{f,g\} \equiv \partial_x f\,\partial_y g - \partial_y f\,\partial_x g$ is the standard Poisson bracket, $\alpha$ is the adiabaticity parameter, $\kappa$ is the normalized background density gradient, and the $\nu_{n,\varpi}(-\nabla_\perp^2)^{N}$ terms model small-scale hyper-dissipation~\cite{Numata2007}.

A field is decomposed into its zonal (flux-surface-averaged) and non-zonal (fluctuating) components via $\langle f\rangle_y = (1/L_y)\!\int\!f\,dy$ and $\tilde f = f - \langle f\rangle_y$. Restricting the resistive coupling to the non-zonal components $(\tilde\phi-\tilde n)$ is the key MHW modification, since the zonal components are constant on a flux surface ($k_\parallel=0$) and therefore do not drive the parallel electron current that the resistive coupling represents~\cite{Numata2007}.

The system admits two quadratic invariants under ideal dynamics:
\begin{align}
    E &= \tfrac{1}{2}\!\int\!\!\big(|\nabla_\perp\phi|^2 + n^2\big)\,dx\,dy,
    \label{eq:E}\\
    W &= \tfrac{1}{2}\!\int\!\!\big(n-\varpi\big)^2 dx\,dy,
    \label{eq:W}
\end{align}
i.e. the total energy $E$ (a sum of kinetic and internal energy) and the generalized enstrophy $W$. The MHW system is linearly unstable for any finite $\kappa$ when numerical dissipation is neglected~\cite{Numata2007}; for the parameters scanned in this work ($\kappa=1$, $\alpha\in[0.1,1]$), the linear dispersion relation gives maximum growth rates $\gamma_{\max}\simeq 0.11-0.15$ at $k_y\rho_s\approx 0.8-1.3$ and $k_x=0$, peaking near $\alpha\approx 0.2-0.3$. Nonlinearly, the MHW system has two asymptotic limits: the hydrodynamic limit ($\alpha\!\to\!0$) where the density and vorticity equations decouple and the dynamics reduces to 2D Navier-Stokes-like turbulence driven by the linear instability, producing nearly isotropic small-scale turbulence, and the adiabatic limit ($\alpha\!\to\!\infty$) where the resistive coupling forces $\tilde n \approx \tilde\phi$, yielding a Hasegawa-Mima-like regime dominated by drift-wave-zonal-flow interaction.

\subsection{Direct numerical simulations}
\label{sec:sims}

Training and evaluation data are generated with the GDB code~\cite{Zhu2018}, which solves Eqs.~\eqref{eq:mhw_n} and \eqref{eq:mhw_w} on a doubly periodic square domain of physical size $(L_x, L_y) = (64,64)$ (in units of $\rho_s$) at a numerical resolution of $(N_x,N_y) = (512,512)$. The spatial discretization uses the Arakawa scheme for the nonlinear Poisson brackets, a pseudospectral treatment for the hyper-diffusion operators (sixth order, $N=3$, with $\nu_n=\nu_\varpi=10^{-5}$ in normalized units), and the trapezoidal leapfrog time integrator with $\Delta t = 2.5\times 10^{-3}$ in normalized time units.

In this study, we fix the normalized background density gradient at $\kappa = 1.0$ and scan the adiabaticity parameter $\alpha$ over $\alpha \in \{0.1,0.2,\dots,1.0\}$ in increments of $0.1$. For each $\alpha$, the simulation is run from random initial perturbations through the linear-growth phase into a statistically stationary saturated state, which is reached within the first $2\times 10^{3}$ normalized time units. Snapshots of $(n,\phi)$ are then recorded at intervals of one time unit throughout the saturated phase, yielding a retained record of $20,000$ snapshots per case. In addition, three simulations at adiabaticity values not used in training, $\alpha=0.25$ and $0.75$ (within the training range) and $\alpha=1.1$ (beyond it), are performed with the same numerical setup and saturation procedure. They serve only as references for the zero-shot tests of Section~\ref{sec:unseen_alpha}, and none of their snapshots enters the training set.

\subsection{Pre-processing and dataset splits}
\label{sec:preproc}
Each $512\times512$ snapshot is downsampled to $128\times128$ by retaining every fourth grid point in each direction, balancing computational efficiency with resolution of energy-containing scales across all $\alpha$ regimes. The density and potential channels are normalized independently to $[-1,1]$ using per-$\alpha$ min-max scaling computed from the training dataset. This normalization is applied consistently during training and inference, and inverted for all post-hoc physical diagnostics. Critically, the train-test split is performed \emph{temporally} rather than randomly: the first 18,500 snapshots from each $\alpha$ case (spanning the initial portion of the saturated phase) form the training set, while the final 1,500 snapshots (from later in the saturated phase) are withheld entirely and reserved for evaluation. This temporal split ensures that training and evaluation cover non-overlapping time intervals, providing a more stringent test of generalization than random splitting from a single time window. Unlike many data-driven approaches that randomly shuffle training and test samples from the same temporal ensemble, our protocol better reflects the challenge of predicting future or unseen transient realizations of turbulent dynamics.

\section{Generative foundation model: \genmodel{} with conditional guidance}
\label{sec:flowmatch}

\subsection{Optimal transport flow-matching framework}
\label{sec:fm}
Flow matching~\cite{Lipman2023} is a generative modeling approach for learning complex data distributions through a continuous normalizing flow. The key idea is to construct a deterministic sample path $\bm{x}(s)$ that evolves from a tractable base distribution (e.g., a standard Gaussian) at pseudo-time $s=0$ to the target data distribution at $s=1$. Here,
$\bm{x}(s)$ represents a spatiotemporal field (in our case, the turbulent density and potential contours), and its evolution is governed by a learned neural velocity field $\bm{v}_\theta$:
\begin{equation}
\frac{d\bm{x}(s)}{ds} = \bm{v}_\theta\!\big(\bm{x}(s), s; \bm{c}_1\big), \quad \bm{x}(0) \sim p_0,
\label{eq:flowode}
\end{equation}
where $p_0$ is a standard Gaussian (the base distribution) and $\bm{c}_1$ is prior conditioning information (e.g., adiabaticity parameter $\alpha$, discussed above). At $s=1$, samples generated by integrating this ODE converge to the data distribution $p_1$.

Training uses optimal-transport (OT) flow matching, which specifies a linear interpolation path between $p_0$ and $p_1$: $$\bm{x}_s = (1-s)\bm{x}_0 + s\bm{x}_1,$$
where $\bm{x}_0 \sim p_0$ and $\bm{x}_1 \sim p_{\text{data}}$. The velocity field is trained to minimize
\begin{equation}
\mathcal{L}_\mathrm{FM}(\theta)
= \mathbb{E}_{s \sim \mathrm{Uniform}[0,1], \, \bm{x}_1 \sim p_{\text{data}}, \, \bm{x}_0 \sim p_0}
\Big\| \bm{v}_\theta\!\big(\bm{x}_s, s; \bm{c}_1\big) - (\bm{x}_1 - \bm{x}_0) \Big\|_2^2.
\label{eq:fmloss}
\end{equation}
Compared to diffusion models, flow matching offers two practical advantages: (i)~the regression target in Eq.~\eqref{eq:fmloss} is simple and simulation-free (no noise schedule or score-matching formulation is needed), and (ii)~the straight-line OT paths are efficient to integrate, typically requiring fewer function evaluations at inference time.
The vector field $\bm{v}_\theta$ is parameterized by a U-Net-style convolutional architecture with sinusoidal embeddings for pseudo-time $s$ and learned embeddings for the conditioning variable $\bm{c}_1$. At inference, samples are generated by solving Eq.~\eqref{eq:flowode} from a random Gaussian seed using a standard ODE integrator (e.g., forward Euler or RK45). The full architectural and training details are provided in Appendix~\ref{sec:architecture}.

\subsection{Multichannel spectral decomposition}
\label{sec:multichannel}

To jointly model the coupled dynamics of the two MHW fields $(n, \phi)$, we adopt the multichannel flow-matching formulation of Parikh~\emph{et al.}~\cite{Parikh2026}. A naive two-channel representation $\bm{x} = [n, \phi]$ would allow the network to learn their joint distribution $p(n, \phi|\alpha)$ and capture cross-correlations between density and potential fluctuations, which is a necessary feature since the MHW dynamics couple $n$ and $\phi$ through the adiabaticity parameter $\alpha$ (Eqs.~\eqref{eq:mhw_n}--\eqref{eq:mhw_w}).
However, this direct two-channel approach suffers from a critical normalization imbalance: the energy-containing scales of $\phi$ (low wavenumbers) dominate the dynamic range of the channel, causing small-scale $\phi$ fluctuations (high wavenumbers) to be numerically suppressed when both fields are normalized independently to $[-1,1]$ during training. This underrepresentation of small-scale potential structure prevents the network from accurately learning the multiscale, coupled physics of the $(n,\phi)$ system.

We address this by decomposing $\phi$ into low and high-wavenumber components before normalization, thereby mitigating the spectral bias inherent in neural networks trained
on normalized data. Define the box wavenumber $q(k_x,k_y) = \max(|k_x|, |k_y|)$ and introduce two overlapping Fourier masks:
\begin{align}
M_{\mathrm{L}}(k_x,k_y) &= \begin{cases} 1 & \text{if } q(k_x,k_y) \le k_h \\ 0 & \text{otherwise} \end{cases}, \\
M_{\mathrm{H}}(k_x,k_y) &= \begin{cases} 1 & \text{if } q(k_x,k_y) \ge k_l \\ 0 & \text{otherwise} \end{cases},
\end{align}
where $(k_l, k_h) = (1.2, 1.6) \, \rho_s^{-1}$ in this study. The corresponding spatial components are
\begin{align}
\phi_{\mathrm{L}} &= \mathcal{F}^{-1}[M_{\mathrm{L}} \hat{\phi}], \\
\phi_{\mathrm{H}} &= \mathcal{F}^{-1}[M_{\mathrm{H}} \hat{\phi}],
\end{align}
where $\hat{\phi} = \mathcal{F}[\phi]$ is the Fourier transform.

Modes satisfying $k_l \le q(k_x,k_y) \le k_h$ appear in both channels. This overlap is essential: it eliminates sharp spectral discontinuities at the cutoff wavenumber $k_h$, which would otherwise introduce training artifacts and impair the network's ability to learn smooth, scale-dependent variations in the dynamics. The three normalized channels $\bm{x} = \left[n, \phi_{\mathrm{L}}, \phi_{\mathrm{H}}\right]$ are then supplied to the flow-matching network. This ensures that small-scale potential fluctuations receive representational weight comparable to large-scale features during training, rather than being drowned out by energy-containing scales.
The effectiveness of this spectral decomposition strategy is validated in Appendix~\ref{sec:ablation_on_spec}, where ablations show that omitting the decomposition results in degraded high-wavenumber fidelity in both unconditional samples and conditional reconstructions across all $\alpha$ regimes, a direct consequence of unmitigated spectral bias.

\subsection{Prior conditioning: classifier-free adiabaticity guidance}
\label{sec:cfguid}

The adiabaticity parameter $\alpha$ encodes the regime of plasma turbulence via the prior conditioning variable ($\bm{c}_1$). During training, $\alpha$ is represented as a learned scalar embedding that is fused with the pseudo-time embedding before being passed to the U-Net backbone.

We employ \emph{classifier-free guidance}~\cite{Ho2022classifier}, a training-time strategy that enables the same network to learn both conditional and unconditional distributions without requiring a separate classifier or likelihood model. Specifically, the $\alpha$ conditioning is randomly removed (i.e., set to zero or a null embedding) with probability $p_{\text{drop}} = 0.2$ during training. This forces the network to learn both
\begin{align}
p(\bm{x}|\alpha) &\quad \text{(conditional distribution at regime $\alpha$)}, \\
p(\bm{x}) &\quad \text{(marginal distribution over all trained $\alpha$ values)}.
\end{align}
Classifier-free guidance offers several practical advantages over classifier-based guidance (which would require training a separate classifier network or loss):
(1) No additional model or likelihood computation is needed; all information flows through the single velocity field $\bm{v}_\theta$.
(2) The network learns to condition on $\alpha$ amortized within a single training objective, avoiding the need for task-specific retraining when switching between conditional and unconditional modes.
(3) Because $\alpha$ is treated as a continuous scalar rather than a discrete class label, the learned conditional distribution can be queried at adiabaticity values not encountered during training, enabling zero-shot interpolation and extrapolation (Section~\ref{sec:unseen_alpha}).

At inference, regime-specific samples are generated by providing the corresponding $\alpha$ value to the network. Unconditional samples are obtained by omitting the $\alpha$ signal, which samples from the learned marginal $p(\bm{x})$.

\subsection{Posterior conditioning: flow-guided measurement assimilation}
\label{sec:posterior}

Inverse problems are naturally formulated as Bayesian posterior inference: given a measurement $\bm{c}_2 = \mathcal{A}(\bm{x}^*)$ acquired under a known state-to-observable operator $\mathcal{A}$, we seek samples from
$$p(\bm{x} \mid \bm{c}_2) \propto p(\bm{c}_2 \mid \bm{x}) \, p(\bm{x}),$$
where $p(\bm{x})$ is our learned prior (conditional or marginal, depending on whether $\alpha$ is known).

We employ a measurement-guided posterior sampling strategy, in the spirit of diffusion posterior sampling (DPS)~\cite{chung2022diffusion}, that combines the learned flow prior with gradient-based guidance~\cite{Parikh2026D-flow}. At each pseudo-time step $s$ along the sampling trajectory, we use Tweedie's estimate to predict the implied data-space target:
\begin{equation}
\hat{\bm{x}}_1(\bm{x}_s) = \bm{x}_s + (1-s) \, \bm{v}_\theta\big(\bm{x}_s, s; \bm{c}_1\big).
\label{eq:tweedie}
\end{equation}
This prediction is then used to compute a measurement consistency loss:
$$J(\bm{x}_s) = \| \bm{c}_2 - \mathcal{A}(\hat{\bm{x}}_1(\bm{x}_s)) \|_2^2.$$
The gradient $\nabla_{\bm{x}_s} J$ is used to steer the sampling ODE towards the posterior:
\begin{equation}
\tilde{\bm{v}}\big(\bm{x}_s, s; \bm{c}_1, \bm{c}_2\big)
= \bm{v}_\theta\big(\bm{x}_s, s; \bm{c}_1\big)
- \gamma \,
\frac{\|\bm{v}_\theta\big(\bm{x}_s, s; \bm{c}_1\big)\|_2}{\|\nabla_{\bm{x}_s} J\|_2 + \epsilon}
\,\nabla_{\bm{x}_s} J,
\label{eq:guided}
\end{equation}
where $\gamma = 0.25$ is a fixed scaling coefficient and $\epsilon = 10^{-8}$ prevents division by zero. The adaptive rescaling in Eq.~\eqref{eq:guided} normalizes the measurement gradient to the magnitude of the model's velocity, ensuring that the likelihood correction remains balanced with the prior drift throughout the sampling trajectory. Without this rescaling, the measurement gradient would either overwhelm the prior dynamics (when large) or be negligible (when small), both pathologies that degrade posterior fidelity~\cite{Parikh2026D-flow}.

The modified ODE is integrated using forward Euler with $N_{\text{steps}} = 200$ uniform steps ($\Delta s = 1/N_{\text{steps}}$). Independent Gaussian seeds yield an ensemble of posterior samples, which we report as ensemble means with per-pixel variance to quantify reconstruction uncertainty.
\emph{No network retraining is required} to apply this posterior sampling strategy to new measurement operators $\mathcal{A}$. The learned velocity field $\bm{v}_\theta$ remains fixed; only the measurement operator and its gradient change. This enables immediate transfer to new inverse problems, a key feature of the proposed foundation model for practical scientific applications where data assimilation problems frequently change.

\subsection{Downstream tasks}
\label{sec:tasks}

The trained flow-matching network, together with the posterior-sampling strategy above, enables three distinct downstream tasks without modification:

\begin{itemize}
    \item \emph{Task 1: Unconditional generation.} Samples are drawn directly from the learned prior by integrating Eq.~\eqref{eq:flowode} with a random seed and no measurement guidance ($\bm{c}_2 = \varnothing$). This produces statistically representative turbulent states that reflect the learned data distribution $p(\bm{x}|\alpha)$ at a specified adiabaticity $\alpha$, or $p(\bm{x})$ when conditioning is omitted. Ensemble statistics of nonlinear functionals of the fields, e.g., the turbulent particle flux (Section~\ref{sec:res_inf}), are evaluated directly on these samples.

    \item \emph{Task 2: Super-resolution.} The measurement operator $\mathcal{A}_{\mathrm{SR}}$ uniformly downsamples the target field by a factor of $r \in \{4, 8\}$: $$\mathcal{A}_{\mathrm{SR}}(\bm{x}) = \bm{x}|_{\text{every } r \text{ gridpoint}}.$$ Given a low-resolution observation $\bm{c}_2 = \mathcal{A}_{\mathrm{SR}}(\bm{x}^*)$, the posterior sampler reconstructs a high-resolution field consistent with this coarse measurement, inferring the fine-scale structure from the learned prior.

    \item \emph{Task 3: State inpainting.} The measurement operator $\mathcal{A}_{\mathrm{IP}}$ applies a binary spatial mask $\mathcal{M}(x,y) \in \{0,1\}^{N_x \times N_y}$: $$\mathcal{A}_{\mathrm{IP}}(\bm{x}) = \mathcal{M} \odot \bm{x},$$ where $\odot$ denotes element-wise multiplication. The posterior sampler then reconstructs the full field from partial spatial observations, inferring unobserved regions from the learned prior while respecting the pixel values where observations are available.
\end{itemize}

Both conditional tasks (Tasks 2 and 3) use the same pre-trained flow-matching backbone without task-specific retraining, realized through the guided sampling scheme of Eq.~\eqref{eq:guided}. This is a core feature of foundation models: a single prior learned on diverse training data can be flexibly reused for downstream tasks with minimal additional computation.

\section{Results}
\label{sec:results}
We organize the results as follows. Section~\ref{sec:res_uncond} establishes, through unconditional sampling, that the model has correctly learned the regime-dependent statistics of the MHW system. Section~\ref{sec:res_inf} shows that statistical transport quantities and their uncertainties can be inferred directly from the generated ensembles. Section~\ref{sec:res_sr} demonstrates physical-state super-resolution from $4\times$ and $8\times$ downsampled fields, and Section~\ref{sec:res_inpaint} full-field reconstruction from spatially partial observations, characterizing how the reconstruction uncertainty is distributed across scales. Section~\ref{sec:unseen_alpha} examines generalization to adiabaticity values not used in training.

All results are evaluated on snapshots withheld from the model's training, as described in Sec.~\ref{sec:preproc}. For unconditional generation, we draw $N=300$ independent samples for each adiabaticity value $\alpha$ and compare their ensemble statistics with $N=300$ held-out DNS snapshots from the corresponding case.

For posterior reconstruction, each measurement is selected from a held-out DNS snapshot. Super-resolution measurements are obtained by uniformly downsampling the $128\times128$ reference field by factors of $4$ and $8$, corresponding to observed resolutions of $32\times32$ and $16\times16$, respectively. Sparse-observation measurements are formed by retaining a prescribed fraction of spatial locations from the reference field. Unless otherwise stated, the same measurement operator is applied to both the density and potential channels. For each fixed measurement, 300 posterior samples are generated from independent Gaussian initializations while keeping the observation unchanged. Throughout the figures, ${n}$ and ${\phi}$ denote the full fluctuating fields modelled by the network, including their zonal components; the zonal/non-zonal decomposition of Section~\ref{sec:mhw} enters only the resistive coupling term.

\subsection{Sampling the learned prior across adiabaticity regimes}
\label{sec:res_uncond}

\begin{figure}[ht!]
    \centering
    \includegraphics[width=0.8\linewidth]{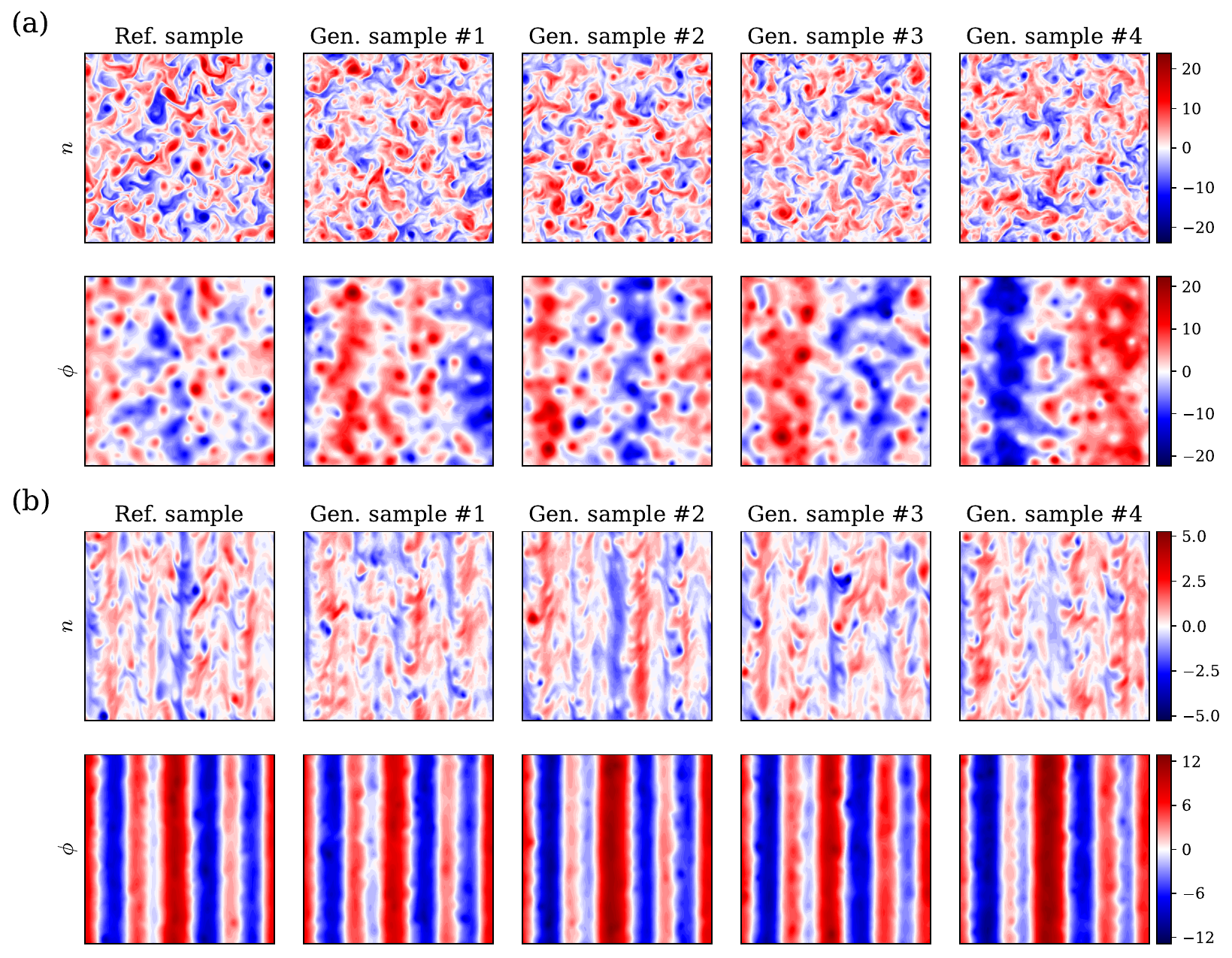}
    \caption{Density ${n}$ (top row) and potential ${\phi}$ (bottom row) contours of a reference (ref.) snapshot and four unconditionally generated (gen.) samples at (a) $\alpha=0.1$ and (b) $\alpha=0.8$. The full doubly periodic $64\times64$ ($\rho_s$ units) domain is shown; the color scale is shared across the panels of each row.}
    \label{fig:uncond_contours}
\end{figure}

We first assess whether the model has captured the regime-dependent structure of the MHW system by drawing samples at each trained value of the adiabaticity parameter $\alpha$ and comparing them against held-out simulation snapshots. The comparison proceeds from qualitative flow structure to quantitative statistics: two-dimensional contours, directional energy spectra, point-wise probability density functions (PDFs), and a distribution-level embedding of the full sample sets. As shown in Fig.~\ref{fig:uncond_contours}, the generated samples reproduce the characteristic transition of the MHW system: at $\alpha=0.1$, both the density ${n}$ and potential ${\phi}$ fields exhibit nearly isotropic small-scale turbulence, whereas at $\alpha=0.8$ the fields are dominated by zonally elongated bands. The generated samples are visually indistinguishable from the reference snapshots, indicating the model successfully synthesizes diverse turbulent states.

\begin{figure}[ht!]
    \centering
    \includegraphics[width=0.95\linewidth]{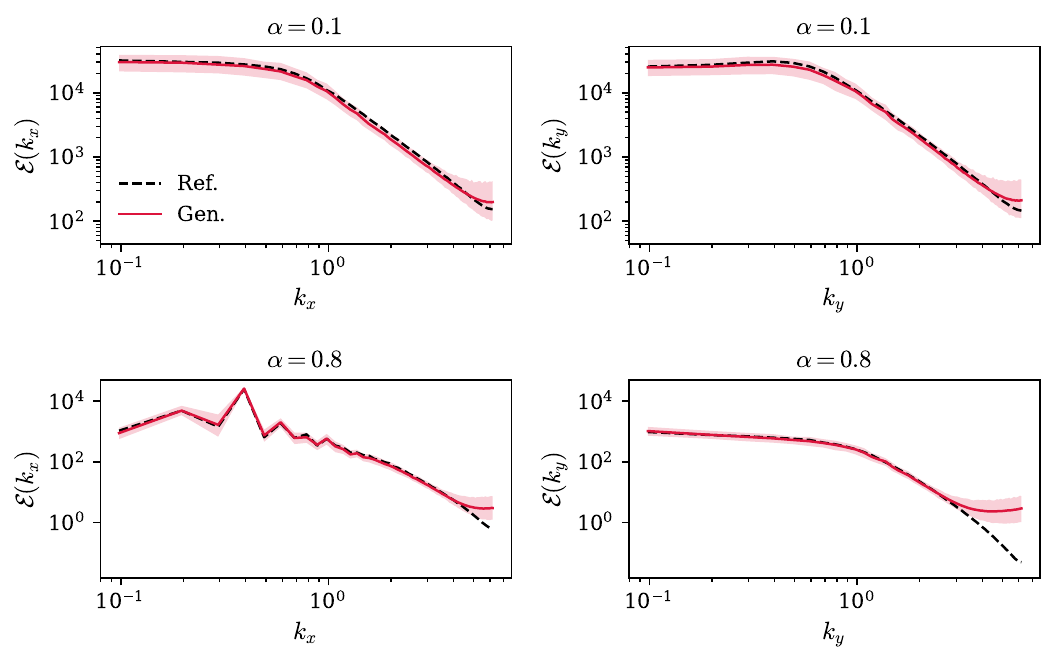}
    \caption{Total energy density spectra  $\mathcal{E}(k_x)$ (left column) and
    $\mathcal{E}(k_y)$ (right column) at $\alpha=0.1$ (top row) and $\alpha=0.8$
    (bottom row). Dashed black lines denote the reference simulation
    statistics; solid red lines denote the mean over unconditionally generated
    samples, with shaded bands indicating the ensemble spread.}
    \label{fig:uncond_spectrum}
\end{figure}

To quantify the fidelity of the generated fields scale by scale, we compare the 1-D total energy density spectra $\mathcal{E}(k_x)$ and $\mathcal{E}(k_y)$ (Eq.~\eqref{eqn:total_energy_density}) between the generated ensemble and the reference statistics at $\alpha=0.1$ and $\alpha=0.8$ (Fig.~\ref{fig:uncond_spectrum}).

At $\alpha=0.1$, the generated spectra follow the reference across the full resolved wavenumber range in both directions, with a slight energy deficit at  wavenumbers ($|k|>5$). At $\alpha=0.8$, the model is able to capture discrete peaks in $\mathcal{E}(k_x)$ associated with the dominant zonal-flow wavenumbers, indicating that the learned prior encodes the nonlinear zonal wavelength selection of the saturated state.
However, the generated spectra flatten at wavenumbers ($k_x > 3, k_y > 2.5$), over-predicting the reference tail. This discrepancy is due to a known spectral bias issue of generative models~\cite{schwarz2021frequency}, which recover low-frequency content preferentially and leave a residual noise floor at small scales. The residual is more visible at $\alpha=0.8$ because the saturated turbulence is roughly two orders of magnitude less energetic than at $\alpha=0.1$, so the high-$k$ reference energy decays to levels comparable with this noise floor. The kinetic and internal energy spectra are shown separately in Appendix~\ref{sec:ablation_on_spec} (Fig.~\ref{fig:spec_baseline_compare}), which also quantifies the benefit of the spectral decomposition for the high-wavenumber floor.

\begin{figure}[h!]
    \centering
    \includegraphics[width=0.9\linewidth]{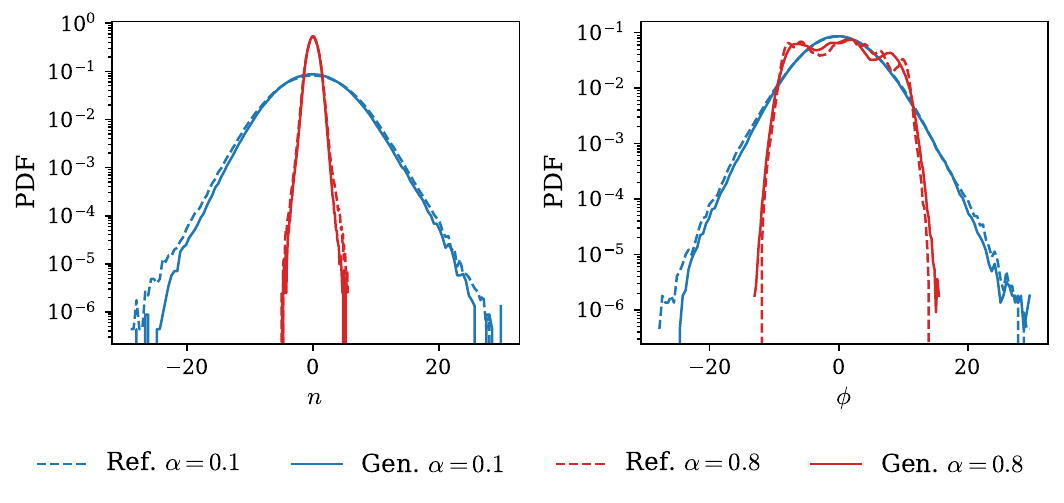}
    \caption{Point-wise PDFs of density ${n}$ (left) and potential
    ${\phi}$ (right) fluctuations at $\alpha=0.1$ (blue) and
    $\alpha=0.8$ (red). Dashed lines denote reference simulation data and
    solid lines denote unconditionally generated samples.}
    \label{fig:uncond_pdf}
\end{figure}

Beyond second-order spectral statistics, we also compare the one-point amplitude statistics of the fluctuations. As shown in Fig.~\ref{fig:uncond_pdf}, the generated PDFs of ${n}$ and ${\phi}$ closely follow the reference in both regimes. Notably, the low-probability tails are recovered down to densities of $\sim\!10^{-5}$ for both cases, with only minor tail deviations in ${\phi}$ at $\alpha=0.8$, confirming that the high-wavenumber noise floor noted above does not distort the amplitude statistics of the fields.

\begin{figure}[h!]
    \centering
    \includegraphics[width=0.8\linewidth]{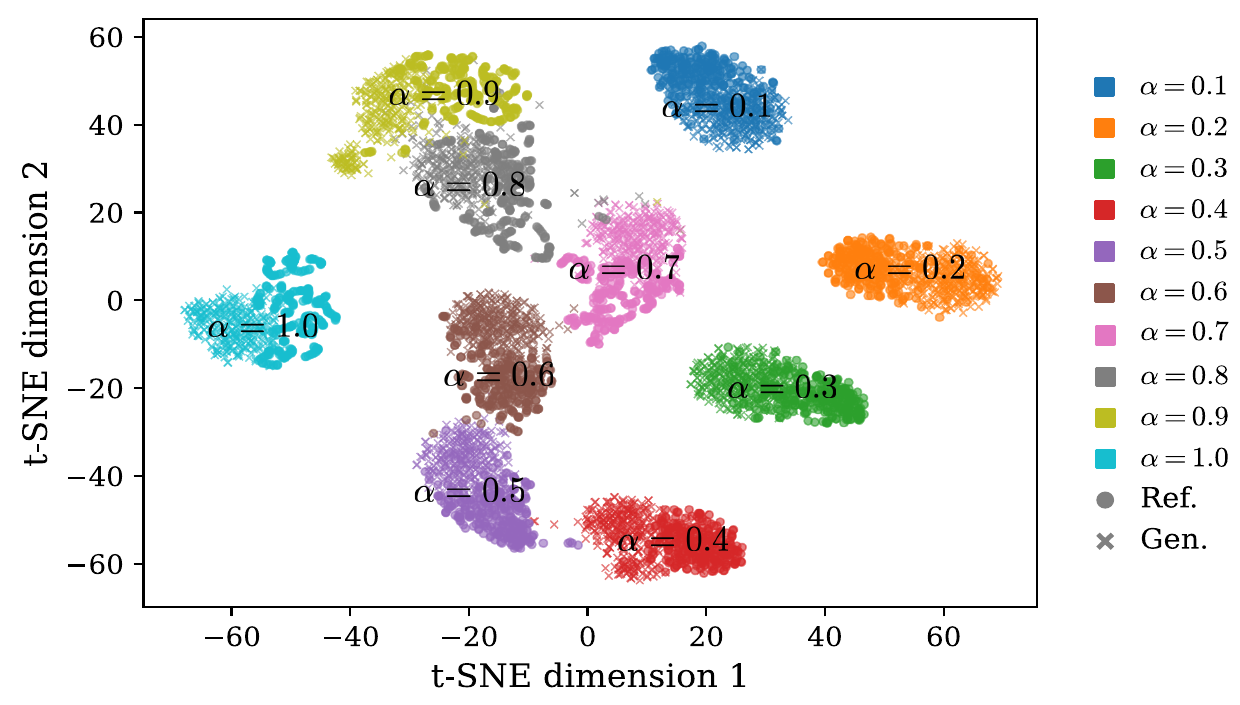}
    \caption{Joint t-SNE embedding of reference (circles) and unconditionally generated (crosses) snapshots for all trained values of $\alpha \in \{0.1, \dots, 1.0\}$.}
    \label{fig:uncond_tsne}
\end{figure}

Finally, we examine whether the generated ensemble covers the reference distribution at the level of the entire dataset, rather than aggregated statistics from individual cases. Figure~\ref{fig:uncond_tsne} shows a joint t-SNE embedding of reference and generated samples across all ten trained values of $\alpha$. The samples organize into well-separated clusters by $\alpha$, and within every cluster the generated samples (crosses) interleave with the reference samples (circles) rather than forming a displaced or contracted sub-cluster. This indicates that conditioning on $\alpha$ reliably steers generation to the correct regime, without sign of either mode collapse or memorization.
The clusters for adjacent $\alpha$ in the strong zonal-flow range ($\alpha=0.7-0.9$) lie closest in the embedding, which is consistent with the underlying physics as the system approaches the adiabatic limit.

\subsection{Inference of statistical transport quantities}
\label{sec:res_inf}
Because the model learns the \emph{joint} distribution of $(n,\phi)$ conditioned on the adiabaticity $\alpha$, statistical transport quantities that are nonlinear functionals of both fields can be evaluated directly on unconditionally generated ensembles, without the model ever having been trained on them. The primary quantity of interest is the radial turbulent particle flux
\begin{equation}
\Gamma_n(\alpha) = \langle \tilde n\, \tilde v_x \rangle_{xy}
              = -\,\langle \tilde n\, \partial_y \tilde\phi \rangle_{xy},
\label{eq:flux}
\end{equation}
where $\tilde v_x = -\partial_y\tilde\phi$ is the radial $E\times B$ velocity and $\langle\cdot\rangle_{xy}$ denotes the average over the full domain, together with the effective turbulent diffusivity $D_\mathrm{eff}(\alpha) = \Gamma_n(\alpha)/\kappa$ obtained from the imposed background gradient; since $\kappa=1$ in this study, $D_\mathrm{eff}$ coincides numerically with $\Gamma_n$ and only $\Gamma_n$ is reported below. Evaluating Eq.~\eqref{eq:flux} over an ensemble of generated snapshots at each $\alpha$ yields not only the mean flux but its full distribution, and hence a data-free estimate of the transport level and its intrinsic variability.

\begin{figure}[h]
    \centering
    \includegraphics[width=0.5\linewidth]{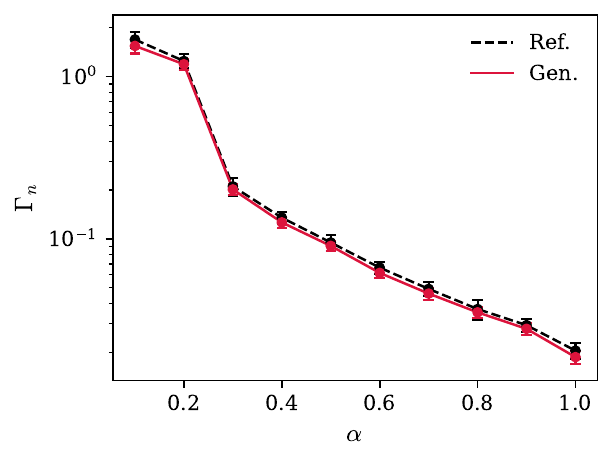}
    \caption{Particle flux versus the adiabaticity parameter $\alpha$ for the ten training cases. The generated ensemble (red) is compared against the DNS reference (black dashed, time-averaged); error bars denote $\pm 1$ standard deviation of the per-snapshot mean flux over each ensemble.}
    \label{fig:flux}
\end{figure}

Figure~\ref{fig:flux} shows the particle flux as a function of $\alpha$ for the ten trained values. The DNS shows a strong reduction in radial transport as $\alpha$ increases. The decrease is particularly sharp between $\alpha=0.2$ and $0.3$, where $\Gamma_n$ drops by nearly one order of magnitude, followed by a more gradual decay toward the adiabatic regime. As the zonal flow becomes dominant, turbulent radial transport is strongly suppressed. The generated ensemble closely follows both the magnitude and the nonlinear dependence of $\Gamma_n$ on $\alpha$, including the sharp transport reduction around $\alpha=0.3$. The predicted mean is lower than the DNS value at all ten values of $\alpha$, by 1 to 12\%, but the discrepancy remains small compared with the overall variation of the flux across regimes.

Furthermore, Table~\ref{tab:stats_quantity} compares the statistical quantities obtained from the DNS and generated ensembles across all trained values of $\alpha$. The model reproduces the kinetic and internal energies (Eqs.~\eqref{eqn: Ek} and~\eqref{eqn:En}), their zonal components (Eqs.~\eqref{eqn:Ekz} and~\eqref{eqn:Enz}), the generalized enstrophy (Eq.~\eqref{eqn:enstrophy}), and the particle flux (Eq.~\eqref{eq:flux}) to within 15\% over the full parameter range, with two exceptions: the enstrophy $W$ is over-predicted at every $\alpha$ (by up to 23\%), consistent with the high-wavenumber noise floor noted in Section~\ref{sec:res_uncond}, and the zonal internal energy $E_{nz}$ at $\alpha=0.1$, where it is only 4\% of $E_n$, is over-predicted by about 70\%. More importantly, it captures the strong redistribution of energy associated with the transition toward the adiabatic regime. As $\alpha$ increases, the zonal kinetic energy becomes an increasingly large fraction of the total kinetic energy, while the internal energy and turbulent particle flux decrease substantially. The ensemble spreads are not reproduced as well as the means: the generated spread is smaller than the DNS spread for $E_n$ and $\Gamma_n$ (by a factor of 0.5 to 0.95) and larger for $E_k$, $E_{kz}$ and $W$ (by up to a factor of 4), so the spread of the generated ensemble should be read as a measure of variability rather than as a calibrated uncertainty. This agreement indicates that the generated fields preserve not only marginal field statistics, but also the regime-dependent energy partition between turbulent fluctuations and self-organized zonal structures.

\begin{table}[h]
\centering
\footnotesize
\setlength{\tabcolsep}{2pt}
\begin{tabular}{|c|c|cccccc|}
\toprule[0.8pt]
$\alpha$ & Model
& $E_k$
& $E_{kz}$
& $E_n$
& $E_{nz}$
& $W$
& $\Gamma_n$ \\
\midrule[0.8pt]

\multirow{2}{*}{$0.1$}
& DNS
& $3.4 \pm 0.206$
& $0.307 \pm 0.0976$
& $11.9 \pm 1.04$
& $0.496 \pm 0.127$
& $26.7 \pm 1.65$
& $1.71 \pm 0.202$ \\
& GenAI
& $3.71 \pm 0.235$
& $0.317 \pm 0.0962$
& $11.1 \pm 0.655$
& $0.859 \pm 0.454$
& $32.2 \pm 5.09$
& $1.55 \pm 0.157$ \\
\midrule[0.8pt]

\multirow{2}{*}{$0.2$}
& DNS
& $3.00 \pm 0.162$
& $0.620 \pm 0.192$
& $6.69 \pm 0.492$
& $0.334 \pm 0.101$
& $18.2 \pm 1.09$
& $1.28 \pm 0.124$ \\
& GenAI
& $3.10 \pm 0.189$
& $0.643 \pm 0.209$
& $6.03 \pm 0.304$
& $0.390 \pm 0.162$
& $20.0 \pm 2.63$
& $1.19 \pm 0.0982$ \\
\midrule[0.8pt]

\multirow{2}{*}{$0.3$}
& DNS
& $2.89 \pm 0.0715$
& $2.45 \pm 0.0909$
& $1.15 \pm 0.129$
& $0.191 \pm 0.0567$
& $3.97 \pm 0.350$
& $0.222 \pm 0.0267$ \\
& GenAI
& $2.73 \pm 0.132$
& $2.32 \pm 0.131$
& $1.01 \pm 0.0623$
& $0.170 \pm 0.0454$
& $4.16 \pm 0.509$
& $0.202 \pm 0.0150$ \\
\midrule[0.8pt]

\multirow{2}{*}{$0.4$}
& DNS
& $3.24 \pm 0.0702$
& $2.92 \pm 0.0762$
& $0.714 \pm 0.0487$
& $0.153 \pm 0.0275$
& $2.66 \pm 0.144$
& $0.136 \pm 0.0112$ \\
& GenAI
& $2.95 \pm 0.178$
& $2.65 \pm 0.179$
& $0.646 \pm 0.0461$
& $0.137 \pm 0.0360$
& $2.78 \pm 0.384$
& $0.126 \pm 0.00975$ \\
\midrule[0.8pt]

\multirow{2}{*}{$0.5$}
& DNS
& $2.18 \pm 0.0626$
& $1.94 \pm 0.0655$
& $0.473 \pm 0.0470$
& $0.0917 \pm 0.0193$
& $2.04 \pm 0.156$
& $0.0984 \pm 0.0104$ \\
& GenAI
& $1.89 \pm 0.0926$
& $1.66 \pm 0.0932$
& $0.432 \pm 0.0267$
& $0.0897 \pm 0.0224$
& $2.26 \pm 0.280$
& $0.0904 \pm 0.00635$ \\
\midrule[0.8pt]

\multirow{2}{*}{$0.6$}
& DNS
& $1.68 \pm 0.0248$
& $1.50 \pm 0.0300$
& $0.351 \pm 0.0283$
& $0.0826 \pm 0.0190$
& $1.53 \pm 0.0850$
& $0.0657 \pm 0.00563$ \\
& GenAI
& $1.65 \pm 0.0686$
& $1.47 \pm 0.0676$
& $0.324 \pm 0.0206$
& $0.0765 \pm 0.0163$
& $1.70 \pm 0.205$
& $0.0616 \pm 0.00431$ \\
\midrule[0.8pt]

\multirow{2}{*}{$0.7$}
& DNS
& $1.74 \pm 0.0292$
& $1.59 \pm 0.0349$
& $0.330 \pm 0.0343$
& $0.108 \pm 0.0236$
& $1.28 \pm 0.0937$
& $0.0489 \pm 0.00492$ \\
& GenAI
& $1.80 \pm 0.0725$
& $1.65 \pm 0.0715$
& $0.334 \pm 0.0262$
& $0.127 \pm 0.0259$
& $1.58 \pm 0.272$
& $0.0460 \pm 0.00403$ \\
\midrule[0.8pt]

\multirow{2}{*}{$0.8$}
& DNS
& $1.61 \pm 0.0274$
& $1.48 \pm 0.0342$
& $0.306 \pm 0.0289$
& $0.122 \pm 0.0220$
& $1.09 \pm 0.100$
& $0.0355 \pm 0.00515$ \\
& GenAI
& $1.60 \pm 0.0634$
& $1.47 \pm 0.0630$
& $0.294 \pm 0.0228$
& $0.121 \pm 0.0222$
& $1.21 \pm 0.167$
& $0.0352 \pm 0.00267$ \\
\midrule[0.8pt]

\multirow{2}{*}{$0.9$}
& DNS
& $1.49 \pm 0.0166$
& $1.37 \pm 0.0197$
& $0.277 \pm 0.0294$
& $0.108 \pm 0.0254$
& $0.996 \pm 0.0633$
& $0.0300 \pm 0.00288$ \\
& GenAI
& $1.47 \pm 0.0696$
& $1.35 \pm 0.0694$
& $0.276 \pm 0.0225$
& $0.121 \pm 0.0225$
& $1.09 \pm 0.146$
& $0.0279 \pm 0.00230$ \\
\midrule[0.8pt]

\multirow{2}{*}{$1.0$}
& DNS
& $1.05 \pm 0.0280$
& $0.964 \pm 0.0306$
& $0.200 \pm 0.0221$
& $0.0741 \pm 0.0142$
& $0.849 \pm 0.0599$
& $0.0212 \pm 0.00222$ \\
& GenAI
& $1.16 \pm 0.0506$
& $1.08 \pm 0.0504$
& $0.194 \pm 0.0171$
& $0.0832 \pm 0.0164$
& $0.861 \pm 0.0986$
& $0.0186 \pm 0.00173$ \\

\bottomrule[0.8pt]
\end{tabular}
\caption{Statistical quantities of the DNS reference and the generative model across $\alpha = 0.1$--$1.0$: kinetic energy $E_k$, zonal kinetic energy $E_{kz}$, internal energy $E_n$, zonal internal energy $E_{nz}$, enstrophy $W$, and particle flux $\Gamma_n$, domain-averaged and reported as mean $\pm$ standard deviation over 300 DNS frames and 300 generated samples per case.}
\label{tab:stats_quantity}
\end{table}

\subsection{Super-resolution from $4\times$ and $8\times$ downsampled observations}
\label{sec:res_sr}
\begin{figure}[h!]
    \centering
    \includegraphics[width=\linewidth]{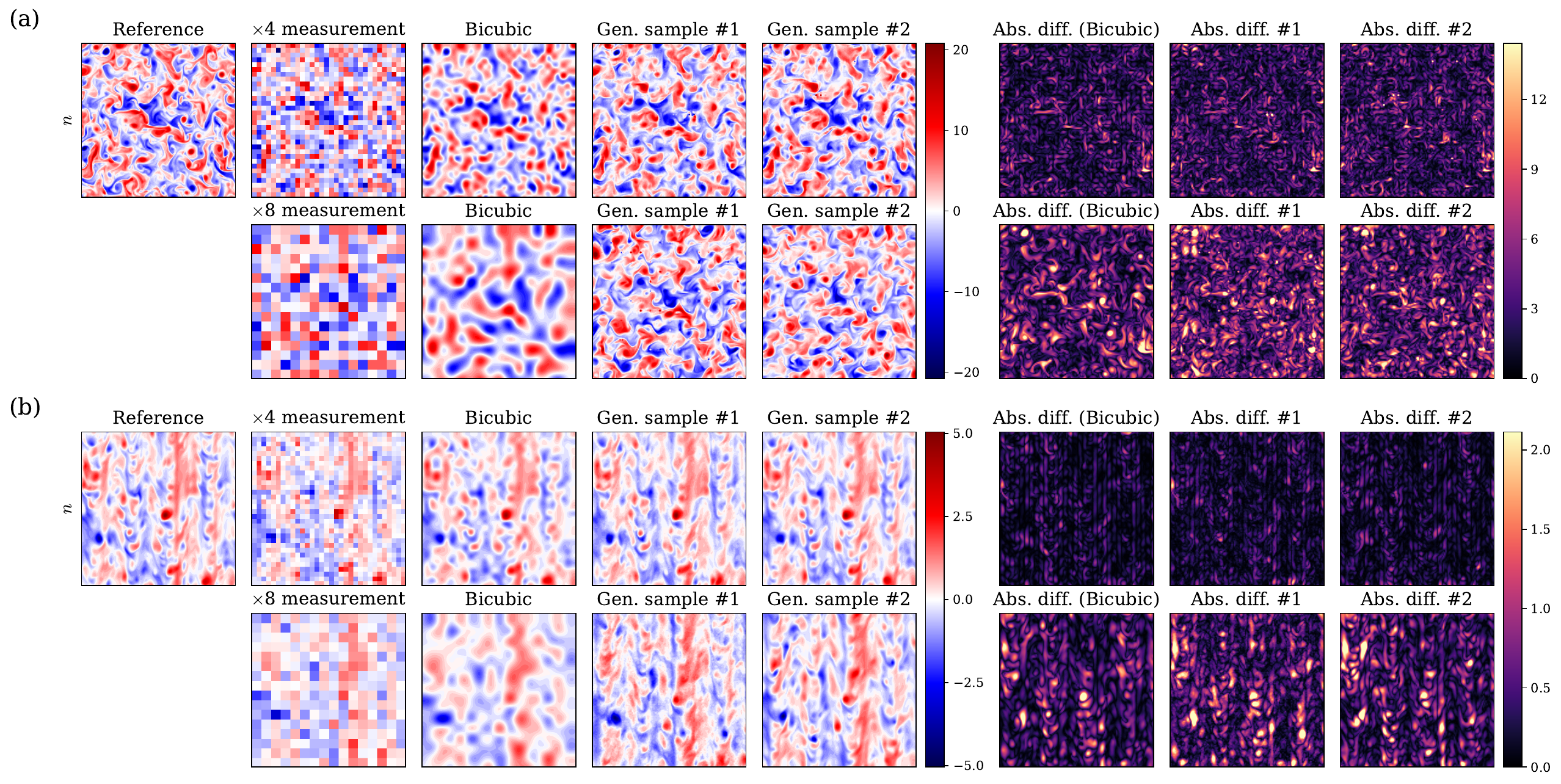}
    \caption{Super-resolution reconstruction contour of density field ${n}$ from $\times$4 and $\times$8 uniformly downsampled measurements at (a)~$\alpha=0.1$ and (b)~$\alpha=0.8$. Left to right: reference, measurement, bicubic interpolation reconstruction, two posterior generated samples, and absolute differences of the bicubic reconstruction and of each posterior sample.}
    \label{fig:fig_dps_contours_den}
\end{figure}

\begin{figure}[h!]
    \centering
    \includegraphics[width=\linewidth]{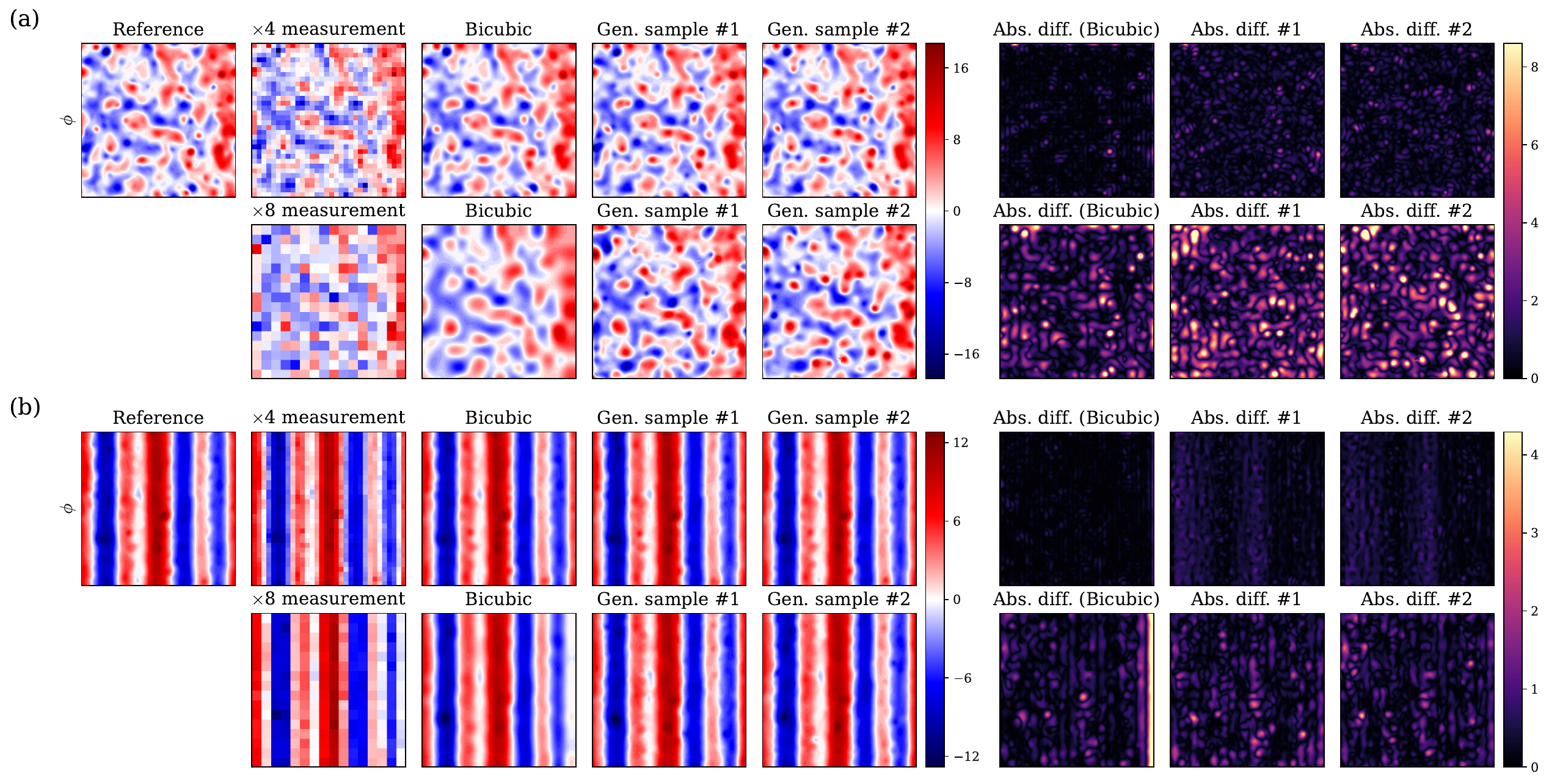}
    \caption{Super-resolution reconstruction contour of potential field ${\phi}$ from $\times$4 and $\times$8 uniformly downsampled measurements at (a)~$\alpha=0.1$ and (b)~$\alpha=0.8$. Left to right: reference, measurement, reconstruction with bicubic interpolation, two posterior generated samples, and absolute differences.}
    \label{fig:fig_dps_contours_phi}
\end{figure}

\begin{figure}[h!]
    \centering
    \includegraphics[width=1\linewidth]{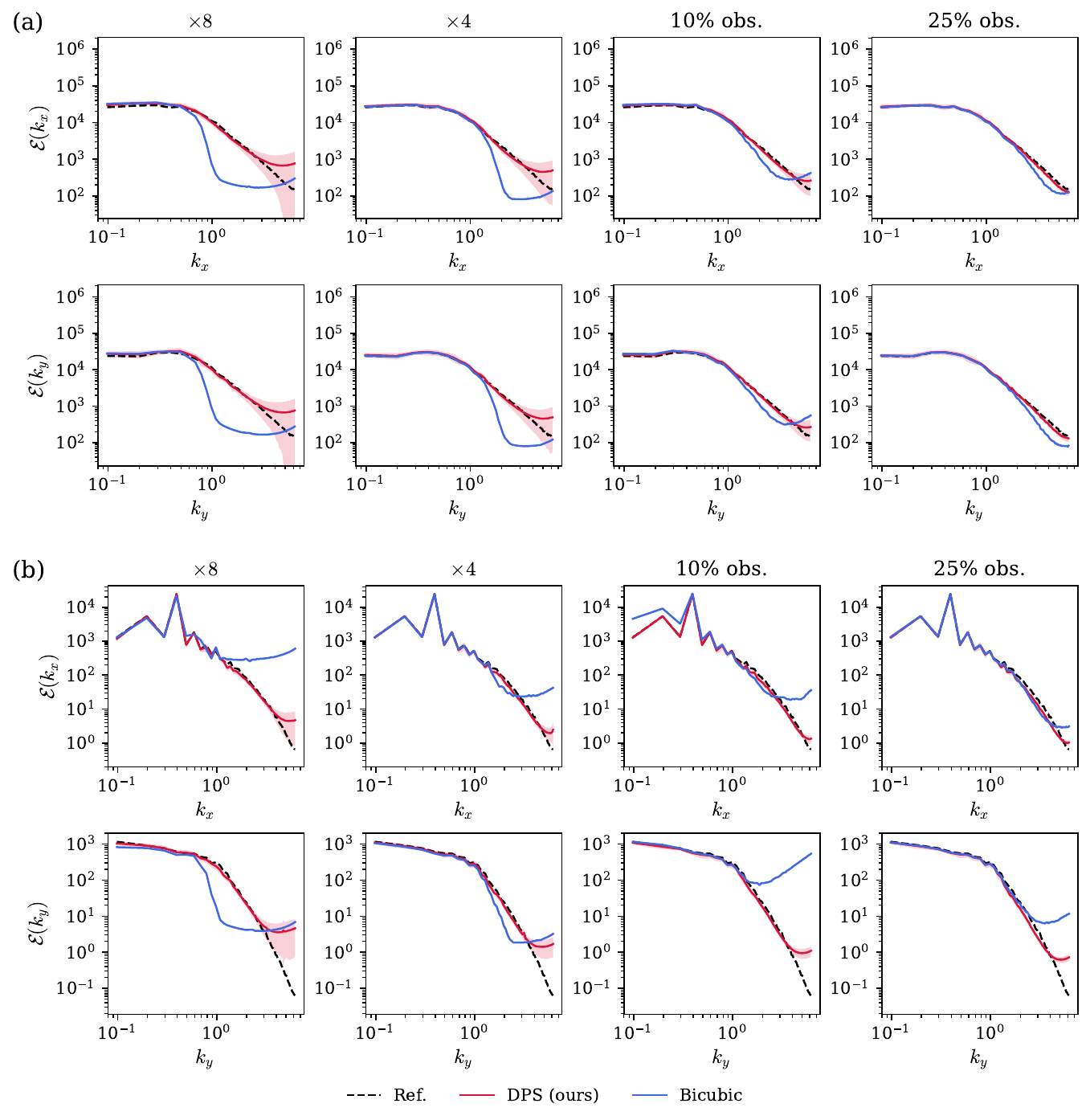}
    \caption{Total energy density spectra $\mathcal{E}(k_x)$ and $\mathcal{E}(k_y)$ of DPS reconstructions (red) versus bicubic interpolation (blue) and the reference (black dashed), for $\times 8$ and $\times 4$ uniformly downsampled measurements and $10\%$ and $25\%$ sparse observations: (a) $\alpha=0.1$, (b) $\alpha=0.8$. }
    \label{fig:fig_dps_spectrum}
\end{figure}

We now impose posterior conditioning, taking $\bm{c}_2$ to be a uniformly decimated observation of the target field that retains every fourth or eighth grid point (effective resolutions of $32\!\times\!32$ and $16\!\times\!16$). Figures~\ref{fig:fig_dps_contours_den} and~\ref{fig:fig_dps_contours_phi} compare the resulting posterior samples with the reference fields and with a deterministic bicubic-interpolation baseline at $\alpha=0.1$ and $\alpha=0.8$; the corresponding energy spectra are shown in the two left columns of Figure~\ref{fig:fig_dps_spectrum}.

The contour comparison (Figs.~\ref{fig:fig_dps_contours_den} and~\ref{fig:fig_dps_contours_phi}) shows that bicubic interpolation recovers only the large-scale organization of the flow, returning overly smooth fields in which the fine-scale eddies are often absent. In contrast, the posterior samples restore small-scale structure consistent with the learned turbulence prior.

For the density field, at $\times 4$ downsampling the observation still constrains most of the energy-containing scales, and the reconstructions are visually nearly indistinguishable from the reference in both regimes, with correspondingly small absolute differences. At $\times 8$ downsampling the reconstructions are no longer pixel-wise exact. This is not a shortcoming of the model but a direct consequence of posterior sampling, since many high-resolution fields are consistent with the same $16\!\times\!16$ observation and individual posterior draws realize different instances of the unresolved scales.

The potential field at $\alpha=0.8$ (Figure~\ref{fig:fig_dps_contours_phi}(b)) is a notable special case in which all reconstructions, including the bicubic baseline, closely match the reference with uniformly small errors; the only conspicuous bicubic artifact is a columnar error along the domain edge at $\times 8$, where non-periodic interpolation is inconsistent with the doubly periodic topology that the learned prior respects automatically. In the near-adiabatic regime $\phi$ is dominated by large-scale zonal structures, so nearly all of its energy resides in modes that survive even $\times 8$ decimation. Super-resolution of $\phi$ is therefore intrinsically easier than that of the density field, whose small-scale content carries a larger share of the energy.

The total energy density spectra in Figure~\ref{fig:fig_dps_spectrum} highlight the distinction between bicubic interpolation and conditional generation quantitatively. The DPS reconstructions follow the reference up to $k\approx 2$--$3$ at both factors, beyond which the same high-wavenumber noise floor as in Section~\ref{sec:res_uncond} appears, most visibly in $\mathcal{E}(k_y)$ at $\alpha=0.8$; bicubic interpolation departs from the reference by one to two orders of magnitude beyond the wavenumber of the coarse grid.

\subsection{Inpainting from partially observed fields}
\label{sec:res_inpaint}
\begin{figure}[h!]
    \centering
    \includegraphics[width=0.9\linewidth]{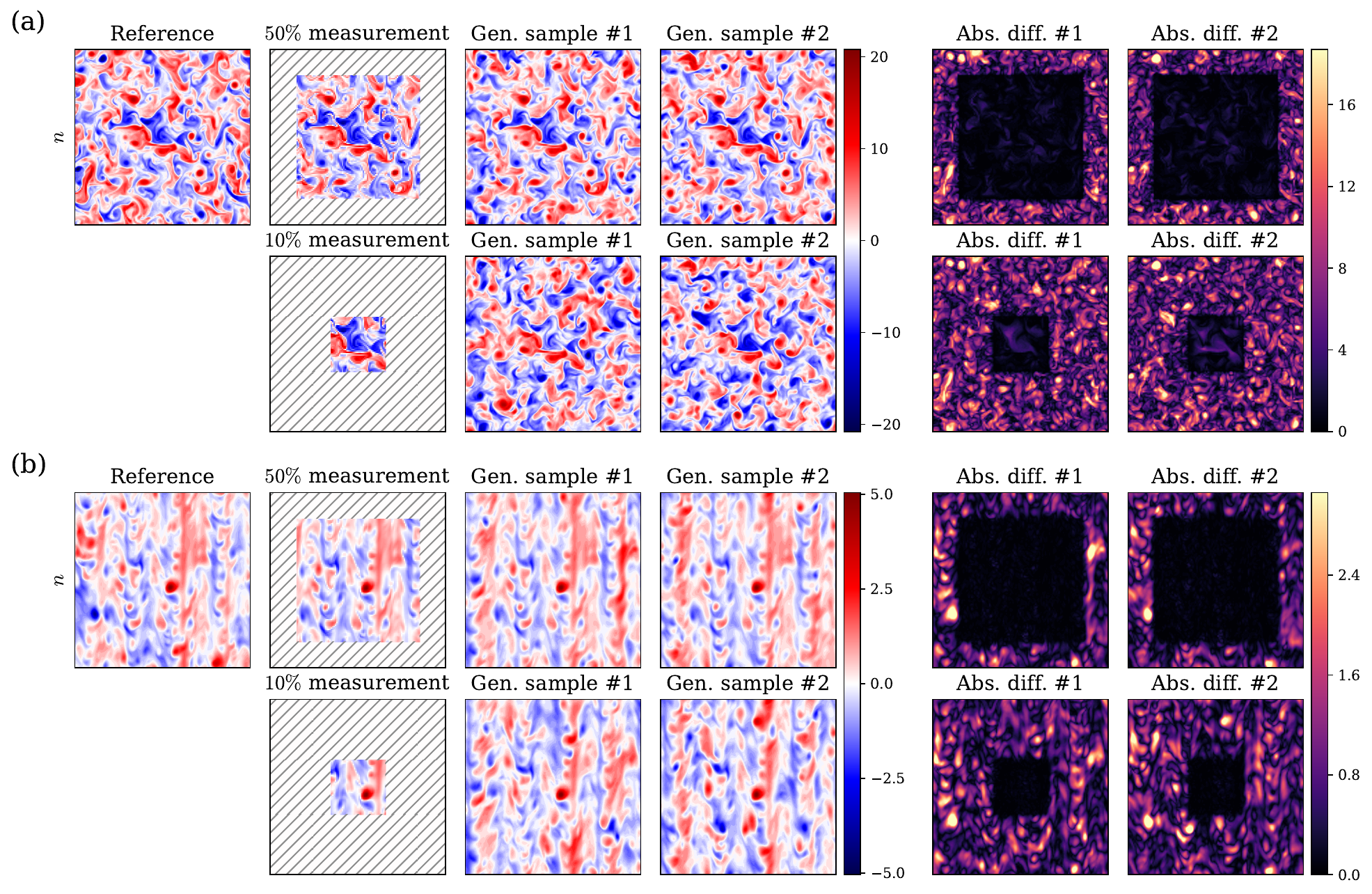}
    \caption{Inpainting contour comparison of ${n}$ from a centered observed region covering 50\% (top) and 10\% (bottom) of the domain, at (a) $\alpha=0.1$ and (b) $\alpha=0.8$. Left to right: reference, measurement, two posterior generated samples, and the absolute difference of each posterior sample. }
    \label{fig:inpaint_den}
\end{figure}

\begin{figure}[h!]
    \centering
    \includegraphics[width=0.9\linewidth]{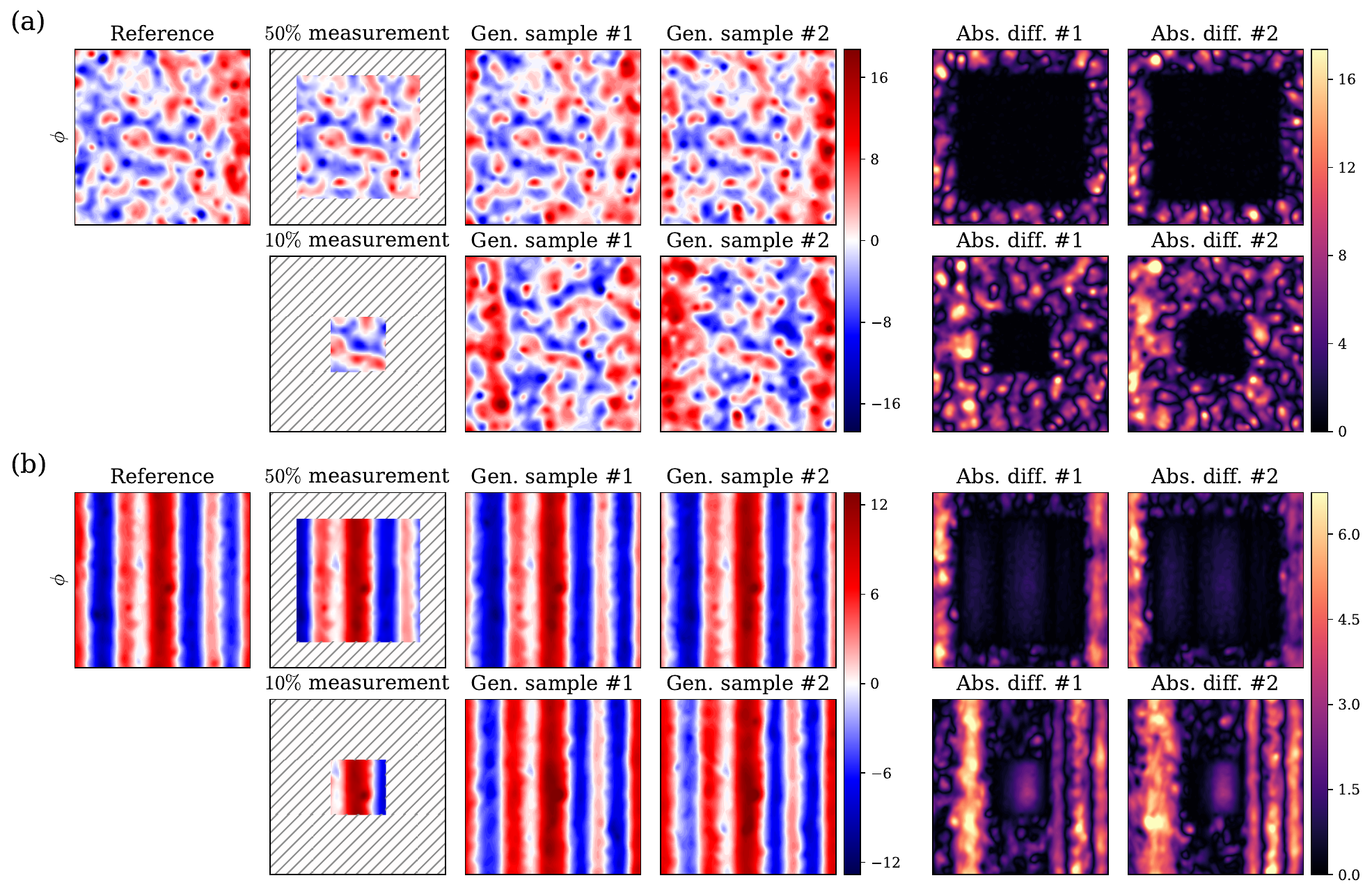}
    \caption{Inpainting contour comparison of ${\phi}$ from a centered observed region covering 50\% (top) and 10\% (bottom) of the domain, at (a) $\alpha=0.1$ and (b) $\alpha=0.8$. Left to right: reference, measurement, two posterior generated samples, and the absolute difference of each posterior sample.}
    \label{fig:inpaint_phi}
\end{figure}

We next consider inpainting, in which $\bm{c}_2$ is a partial observation of the field, either a contiguous observation window or randomly placed points, and the model completes the surrounding flow. Figures~\ref{fig:inpaint_den} and~\ref{fig:inpaint_phi} show reconstructions of the density and potential fields from a centered square region covering $50\%$ and $10\%$ of the domain at $\alpha=0.1$ and $\alpha=0.8$.

Within the observed window the posterior samples reproduce the measurement faithfully, and only small amplitude residuals remain in the absolute-difference panels. These residuals reflect the soft nature of the conditioning in Eq.~\eqref{eq:guided}, where measurement consistency is imposed through the gradient of the misfit rather than by exact replacement of the observed pixels, so the observations are matched only up to the tolerance of the guided dynamics. Outside the window the samples depart substantially from the reference realization, because the turbulence decorrelates within a few structure sizes and the unobserved region admits many valid realizations. Crucially, the generated structures connect to the observed region without visible discontinuity at the window boundary. Each sample is therefore a physically plausible realization consistent with the measurement.

\begin{figure}[H]
    \centering
    \includegraphics[width=0.85\linewidth]{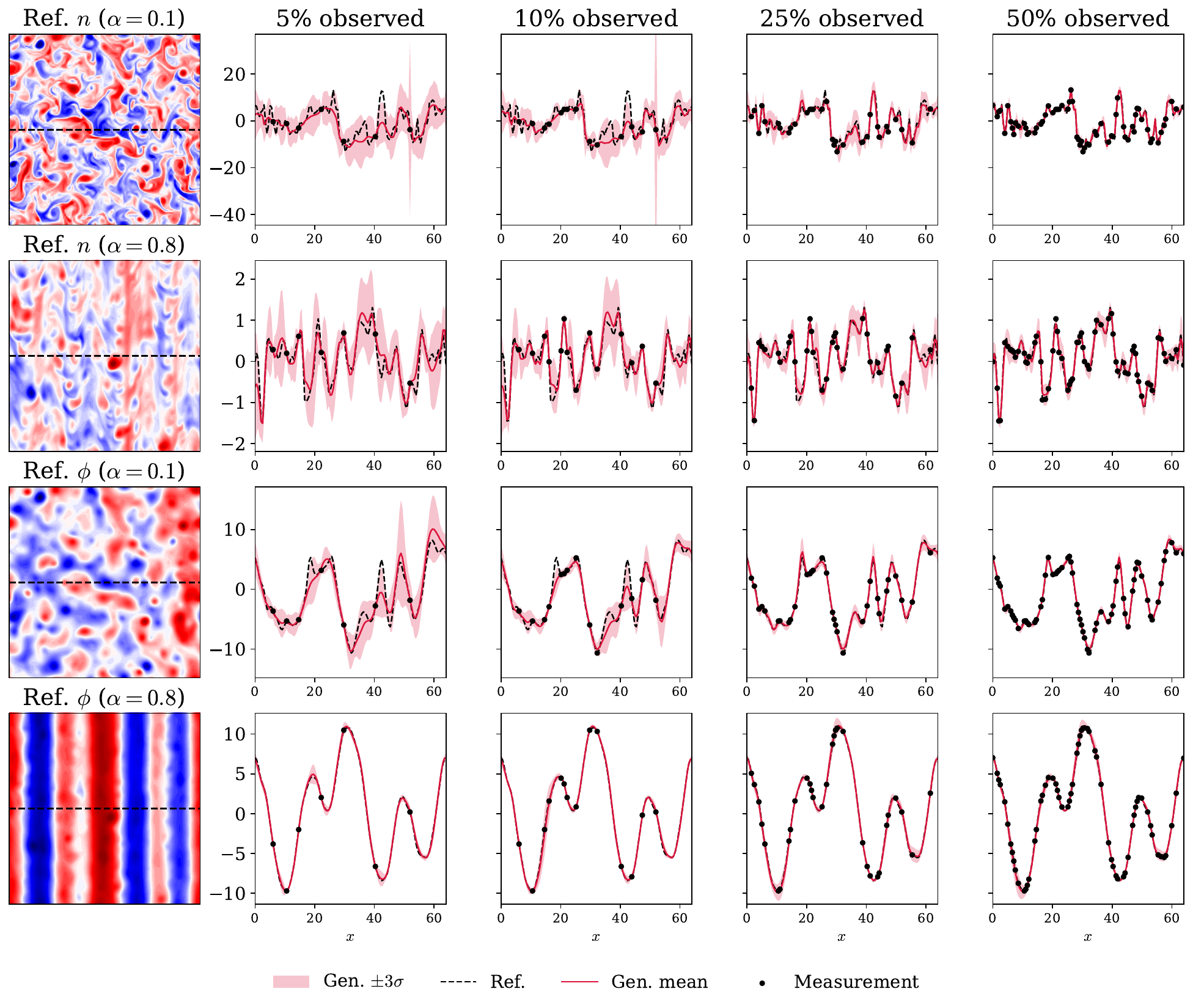}
    \caption{Posterior uncertainty of DPS reconstructions from randomly observed pixels along a horizontal line cut. Each row shows a reference  snapshot followed by reconstructions from 5\%, 10\%, 25\%, and 50\% randomly observed pixels.}
    \label{fig:dps_random_uq}
\end{figure}

Figure~\ref{fig:dps_random_uq} extends the analysis to randomly placed point observations and quantifies the posterior uncertainty along a horizontal line cut, with the ensemble mean and a $\pm3\sigma$ band (pixel-wise standard deviation $\sigma$) computed from repeated posterior draws at observation fractions of $5\%$, $10\%$, $25\%$, and $50\%$. First, the uncertainty band contracts monotonically as the observation fraction increases, for both fields and both regimes, as the posterior concentrates around the true realization.
Second, at $\alpha=0.8$ the uncertainty remains low wherever zonal structures dominate, even at the sparsest observation level. Because these coherent structures possess long correlation lengths, a few observed pixels suffice to pin them across the entire domain, whereas the fine-scale eddies of the hydrodynamic regime remain uncertain between sensors. The right two columns of Figure~\ref{fig:fig_dps_spectrum} confirm that reconstructions from $10\%$ and $25\%$ random observations likewise recover the reference energy spectra up to $k\approx 2$--$3$, with the same high-wavenumber floor, closing the comparison presented in Section~\ref{sec:res_sr}.

\subsection{Zero-shot predictions on unseen $\alpha$}
\label{sec:unseen_alpha}
\begin{figure}[h]
    \centering
    \includegraphics[width=0.85\linewidth]{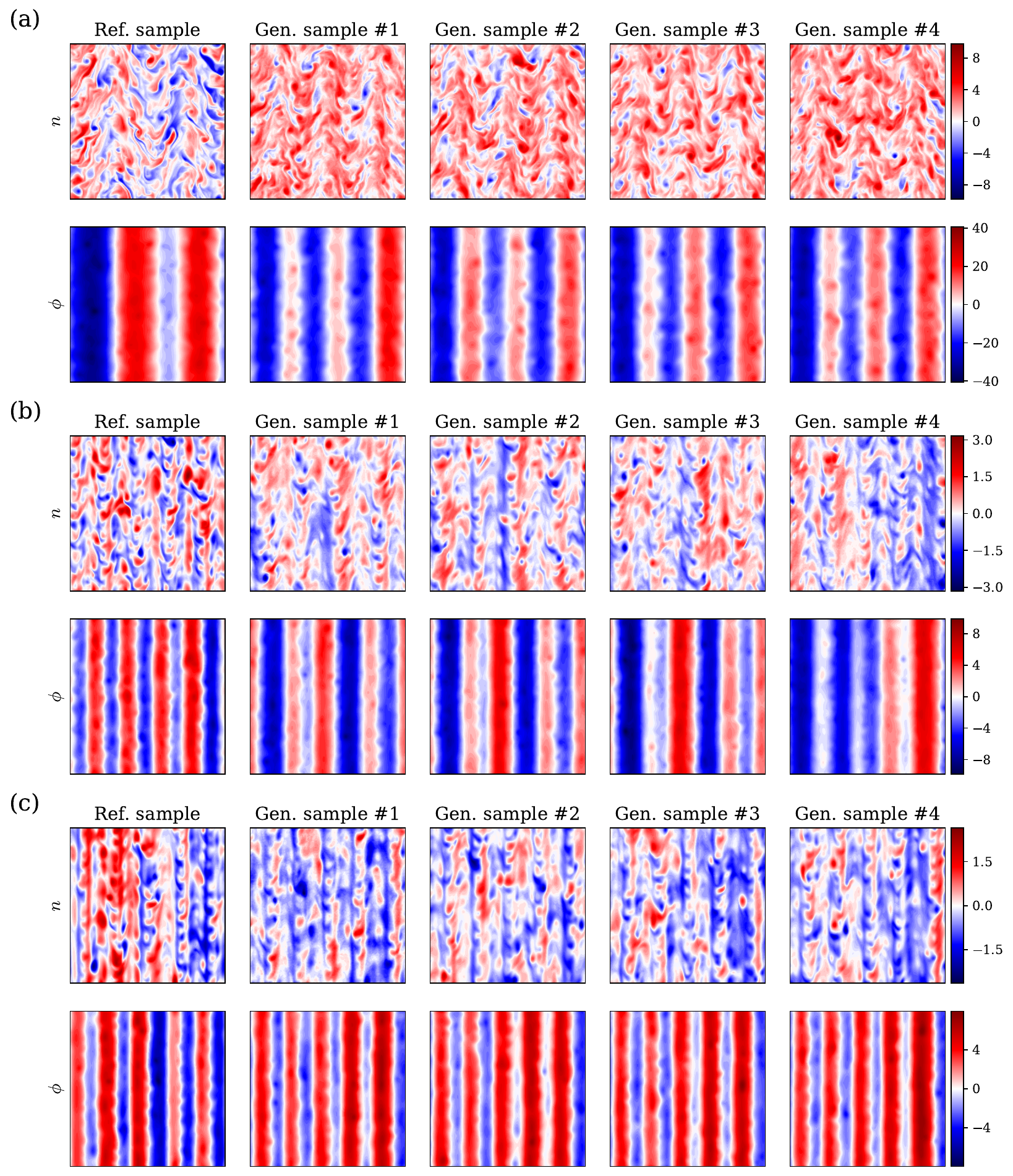}
    \caption{Zero-shot unconditional generation at unseen adiabaticity values $\alpha=0.25$, $0.75$, and $1.1$. The cases $\alpha=0.25$ and $0.75$ lie within the training range, while $\alpha=1.1$ lies outside the training range.}
    \label{fig:uncond_new_alpha_contours}
\end{figure}

We next examine whether the learned model can generalize to adiabaticity values that are not included during training. We consider $\alpha=0.25$ and $0.75$ for interpolation within the training range, and $\alpha=1.1$ for extrapolation beyond the largest training value. During inference, these values are supplied directly through the continuous $\alpha$ embedding, without retraining or fine-tuning. Figure~\ref{fig:uncond_new_alpha_contours} shows that prior samples at all three unseen values reproduce the main regime-dependent flow morphology observed in the corresponding DNS. In particular, the model captures the change from more disordered turbulence at $\alpha=0.25$ to increasingly pronounced zonal organization at $\alpha=0.75$ and $1.1$. This indicates that the learned conditioning is not restricted to the discrete $\alpha$ values encountered during training.

\begin{figure}[ht]
    \centering
    \includegraphics[width=1\linewidth]{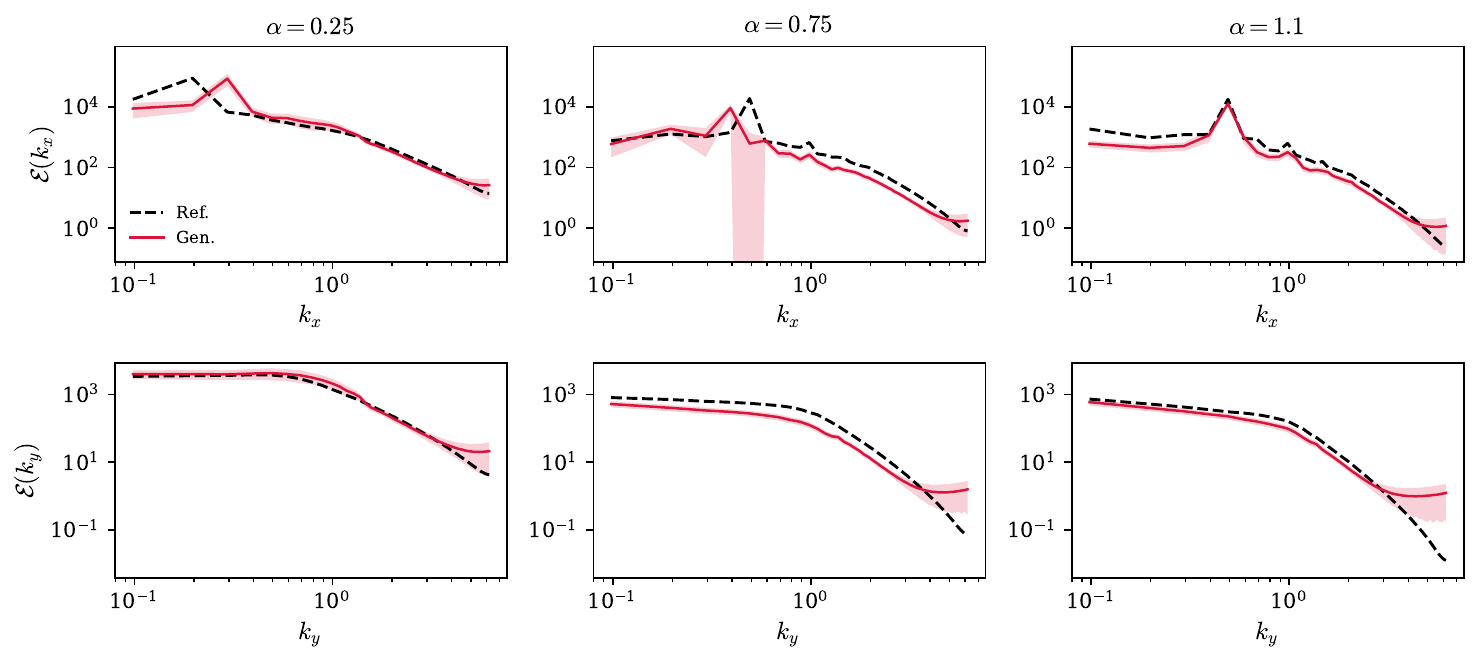}
    \caption{Total energy density spectra $\mathcal{E}(k_x)$ (top row) and $\mathcal{E}(k_y)$ (bottom row) at unseen adiabaticity values $\alpha=0.25$, $0.75$, and $1.1$. Dashed black lines denote the reference simulation statistics; solid red lines denote the mean over unconditionally generated samples, with shaded bands indicating the ensemble spread.}
    \label{fig:uncond_new_alpha_spectrum}
\end{figure}

However, the agreement is weaker when the generated ensembles are compared through their total energy spectral density. As shown in Fig.~\ref{fig:uncond_new_alpha_spectrum}, the model reproduces the overall spectral shape and the dominant zonal peaks, but noticeable errors remain at the lowest $k_x$ modes for all three unseen values. These modes contain the largest-scale structures of the flow and are strongly dependent on $\alpha$. Although the training dataset contains many turbulent realizations at each condition, the dependence on $\alpha$ is sampled at only ten discrete values. The scalar $\alpha$ label therefore provides relatively insufficient information for identifying the precise statistical distribution at an unseen condition. Interpolation of the conditioning embedding can recover the qualitative regime and dominant structures, but does not determine the correct energy partition among the lowest wavenumbers.

\begin{figure}[H]
    \centering
    \includegraphics[width=0.8\linewidth]{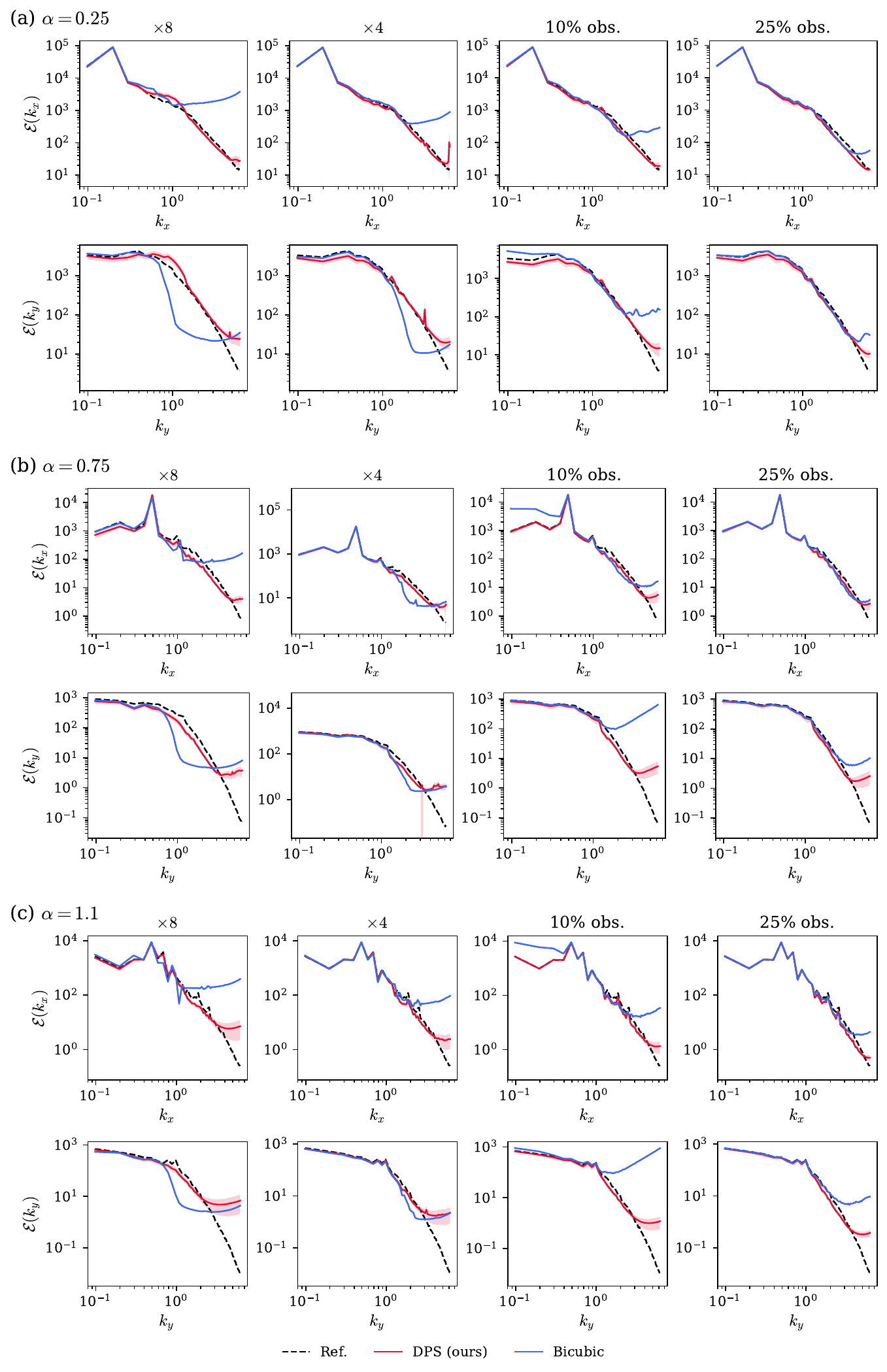}
    \caption{Total energy density spectra $\mathcal{E}(k_x)$ and $\mathcal{E}(k_y)$ of DPS reconstructions (red) versus bicubic interpolation (blue) and the reference (black dashed), for $\times 8$ and $\times 4$ uniformly downsampled measurements and $10\%$ and $25\%$ sparse observations at unseen adiabaticity values: (a) $\alpha=0.25$, (b) $\alpha=0.75$, and (c) $\alpha=1.1$.}
    \label{fig:dps_new_alpha_spectrum}
\end{figure}

The situation changes substantially once partial field measurements are introduced through DPS. Figure~\ref{fig:dps_new_alpha_spectrum} compares the reconstructed spectra with bicubic interpolation for the same three unseen values of $\alpha$. Across both super-resolution and sparse-observation settings, the DPS reconstructions recover the reference spectra more faithfully than the bicubic baseline. Specifically, the improvement is clear beyond the scales directly resolved by the measurements, where bicubic interpolation flattens while the generative reconstruction retains the turbulent spectral decay up to $k\approx 2$--$3$, beyond which the high-wavenumber floor noted above sets in.

Notably, small localized spectral spikes are visible in some super-resolution cases, most clearly near $k_y\approx3$ for the $\alpha=0.25$ case. These features are associated with the discrete measurement constraint rather than with a physical turbulent mode. Uniform decimation constrains the reconstruction only on a regularly spaced sampling lattice, while the unresolved values between measurement points remain inferred from the learned prior. Strong DPS guidance toward exact consistency on this lattice can therefore introduce a weak grid-scale imprint in the reconstructed field, which appears as a narrow spectral excess at intermediate wavenumbers. The artifact is small relative to the broadband spectra and decreases as the observations provide stronger constraints on the unresolved field.

The improvement brought by the measurements highlights the complementary roles of parameter conditioning and physical observations. The scalar $\alpha$ label identifies the approximate turbulence regime, but by itself does not specify which region of the conditional state distribution should be sampled. The measurements provide realization-specific information and substantially reduce this ambiguity. Therefore, they constrain the posterior toward the part of the learned manifold that is consistent with the actual flow state, even when the supplied $\alpha$ was not encountered during training. As the observation density increases, the posterior becomes more strongly constrained and the reconstructed spectra approach the DNS reference. This trend is visible when moving from $\times8$ to $\times4$ super-resolution and from $10\%$ to $25\%$ spatial observations.

\section{Summary and outlook}
\label{sec:summary}

We have presented a proof-of-principle study of generative AI applied
to magnetized plasma turbulence, in which a single multichannel
flow-matching model with a spectral decomposition strategy was trained on
2D \MHW{} simulations spanning the full hydrodynamic-to-adiabatic
parameter range. The trained model serves a triple role: it
unconditionally generates statistically faithful quasi-steady-state
snapshots that reproduce regime-dependent energy spectra and zonal-flow
morphology, from which the turbulent particle flux and the regime-dependent
energy partition are inferred directly; it performs physical-state super-resolution from
heavily downsampled fields ($4\times$ and $8\times$); and it
reconstructs the full $(n,\phi)$ fields from spatially sparse
observations down to $\sim\!10\%$ coverage. The learned conditioning
also extends to adiabaticity values not used in training, qualitatively
for unconditional samples and quantitatively once measurements are
supplied. To our knowledge this is
the first demonstration that a conditional generative foundation model
can simultaneously capture small-scale drift-wave turbulence and the
coexisting self-generated zonal flow within a single learned prior,
and provide both reconstruction and uncertainty estimates without
task-specific retraining.

Several limitations of the present work, and corresponding directions
for follow-up, are worth highlighting. First, the present study is
restricted to a single 2D fluid model in a small periodic box and to
quasi-steady-state snapshots; extension to (a) the original HW
model and other reduced models, (b) 3D and gyrofluid systems, and
(c) the transient and bursty regimes, e.g., LH-like and ELM-like dynamics,
are natural next steps. Second, our reconstruction uses static
snapshots; conditioning on temporal context, or jointly modelling the
spatio-temporal dynamics, would likely improve the reconstruction of the
small-scale turbulence under sparse observations and enable a true
\emph{forecasting} mode of operation. Third, the model is trained on numerical data with known
physics; testing on experimental measurements, e.g., 2D imaging
diagnostics on tokamaks or stellarators, and comparison against
existing reduced-order and operator-learning approaches will be necessary to
establish utility for routine plasma diagnostics. Fourth, a
particularly attractive target in this direction is \emph{cross-field}
inference. Unlike the PINN-based approach~\cite{mathews2021uncovering}, which requires knowledge of the underlying governing equations or an approximation of them, in our study, because the learned prior is joint over $(n,\phi)$, observations of the density channel alone, the quantity most readily
imaged experimentally, can in principle constrain posterior samples
of the unobserved potential, and hence of the $E\times B$ flow and the
turbulent flux; quantifying the fidelity of such density-to-potential
reconstruction is a natural next experiment within the present
framework. Finally, under extreme observational sparsity the fine-scale content
remains under-determined between sensors, whereas the long-correlation
zonal component is pinned by even a few observations. Recovering this small-scale content
in the sparse limit may benefit from temporal context or
physics-informed conditioning, for instance through constraints derived
from the conservation laws.

Taken together, our results suggest that flow-matching-based generative
models offer a flexible, regime-agnostic foundation for plasma-turbulence
diagnostics, sparse-data reconstruction, and statistical inference, and
constitute a promising step toward generative \emph{foundation} models
for the broader class of self-organizing magnetized plasma systems.

\begin{acknowledgments}
This work was supported by the U.S. Department of Energy (DOE) Genesis Mission under DE-SC0026662. This research used resources of the National Energy Research Scientific Computing Center, a DOE Office of Science User Facility supported by the Office of Science of the U.S. DOE under Contract No. DE-AC02-05CH11231 (FES-ERCAP 0036750 and DDR-ERCAP0039129). The authors would also like to acknowledge the support from the NVIDIA Academic Grant Program with the use of NVIDIA RTX PRO 6000 GPUs and NVIDIA CUDA Toolkit.
\end{acknowledgments}

\section*{Data Availability Statement}
The data that support the findings of this study are available from the corresponding author upon reasonable request.

\appendix
\section{Definitions of physical quantities}
\label{sec:definition}
We summarize here the physical diagnostics used throughout the quantitative evaluation. In Fourier space, the potential, kinetic, and total energy densities are defined as
\begin{align}
\mathcal{E}_n(k_x,k_y) &= \frac{1}{2}|\hat n(k_x,k_y)|^2, \label{eqn:potential_energy_density} \\
\mathcal{E}_k(k_x,k_y) &= \frac{1}{2}(k_x^2+k_y^2)|\hat\phi(k_x,k_y)|^2 , \label{eqn:kinetic_energy_density}\\
\mathcal{E}(k_x,k_y) &= \mathcal{E}_n(k_x,k_y) + \mathcal{E}_k(k_x,k_y) . \label{eqn:total_energy_density}
\end{align}
The corresponding domain-averaged quantities are
\begin{align}
E_k    &= \frac{1}{2L_xL_y}\int \left|\nabla_\perp \phi\right|^2\,dx\,dy, \label{eqn: Ek}\\
E_{kz} &= \frac{1}{2L_xL_y}\int \left(\partial_x \langle \phi \rangle_y\right)^2\,dx\,dy, \label{eqn:Ekz}\\
E_n    &= \frac{1}{2L_xL_y}\int n^2\,dx\,dy, \label{eqn:En}\\
E_{nz} &= \frac{1}{2L_xL_y}\int \langle n \rangle_y^2\,dx\,dy, \label{eqn:Enz}
\end{align}
where $E_{kz}$ and $E_{nz}$ quantify the kinetic and density energy contained in the zonal components. We further evaluate the generalized enstrophy,
\begin{equation}
W = \frac{1}{2L_xL_y}\int \left(n-\nabla_\perp^2\phi\right)^2\,dx\,dy, \label{eqn:enstrophy}
\end{equation}
together with the radial turbulent particle flux defined in Eq.~\eqref{eq:flux}. These diagnostics are computed for both DNS and generated ensembles and are used to assess the fidelity of the learned joint distribution beyond point-wise field statistics.

\section{The effectiveness of spectral decomposition}
\label{sec:ablation_on_spec}
\begin{figure}[h]
    \centering
    \includegraphics[width=1\linewidth]{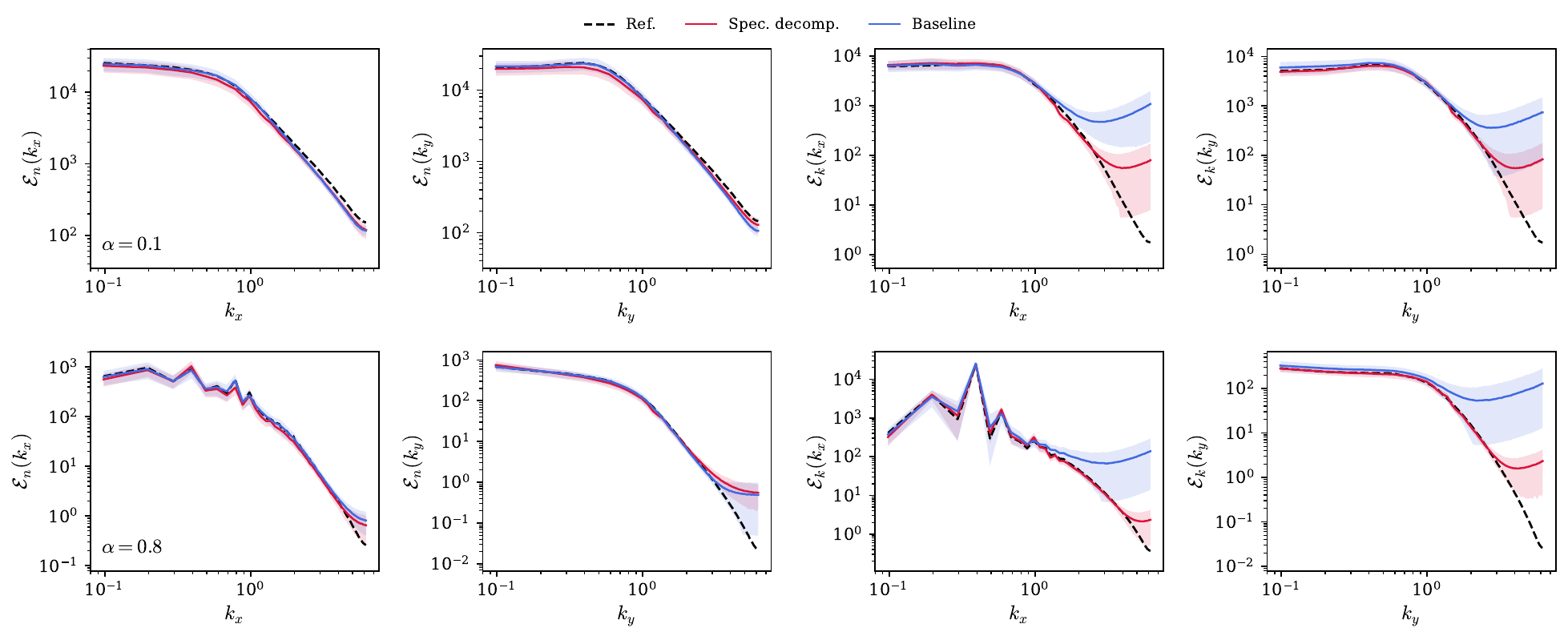}
    \caption{Spectra of unconditional samples with and without spectral decomposition. Columns show the density spectra $\mathcal{E}_n(k_x)$, $\mathcal{E}_n(k_y)$ and kinetic-energy spectra $\mathcal{E}_k(k_x)$, $\mathcal{E}_k(k_y)$; rows correspond to $\alpha=0.1$ (top) and $\alpha=0.8$ (bottom). Each panel compares the reference statistics with the sample mean $\pm$ one standard deviation from the spectral-decomposition model (red) and the baseline model trained without spectral-decomposition (blue).}
    \label{fig:spec_baseline_compare}
\end{figure}

In this section, we evaluate the statistical fidelity of unconditionally generated states and isolate the contribution of the spectral-decomposition strategy introduced in Section~\ref{sec:flowmatch}. For each of the ten trained adiabaticity values $\alpha = 0.1, \ldots, 1.0$ we compare the spectra following the same protocol as in Section~\ref{sec:res_uncond}. As shown in Fig.~\ref{fig:spec_baseline_compare}, both models reproduce the potential energy density spectra $\mathcal{E}_n(k_x)$ and $\mathcal{E}_n(k_y)$ over the entire resolved range and agree with the reference despite minor deviations after $k_y\approx 2$. However, the kinetic spectra $\mathcal{E}_k$ from the baseline model flatten beyond $k_x,k_y \approx 2$, whereas the spectral decomposition model shows decent alignment with the reference for both $\alpha = 0.1$ and $\alpha = 0.8$. The improvement is confined to $E_k$ because only the potential is split into low- and high-wavenumber channels, and the steep decay of $|\hat{\phi}|^{2}$ leaves its small scales near the normalization floor where the $k^{2}$ weighting of $\mathcal{E}_k$ amplifies any residual error.

\section{Network architecture and training details}
\label{sec:architecture}

The flow-matching vector field is parameterized by a U-Net with a base channel width of 128 and channel multipliers $[1,1,2,3,4]$ across five resolution levels. Each level contains two residual blocks, and multi-head self-attention with eight heads is applied at spatial resolutions $8$, $16$, and $32$. The model uses a noise scale of $\sigma=10^{-3}$.

Training uses the Adam optimizer with an initial learning rate of $2\times10^{-4}$ and a cosine-annealing schedule over 40 epochs. The batch size is 40, distributed across eight GPUs. At inference, the probability-flow ODE is integrated using 200 forward-Euler steps.

\bibliography{cgen}
\end{document}